\documentclass[traditabstract]{aa}

\usepackage[hyperindex,breaklinks=true, colorlinks, citecolor=blue]{hyperref}  % for active links 

\newcommand{\postref}[1]{{\bf #1}}
\renewcommand{\postref}[1]{#1}

\def\deg{\ifmmode^\circ\else$^\circ$\fi}
\def\pdeg{\ifmmode $\setbox0=\hbox{$^{\circ}$}\rlap{\hskip.11\wd0 .}$^{\circ} \else \setbox0=\hbox{$^{\circ}$}\rlap{\hskip.11\wd0 .}$^{\circ}$\fi}
\def\arcs{\ifmmode {^{\scriptstyle\prime\prime}} \else $^{\scriptstyle\prime\prime}$\fi}
\def\arcm{\ifmmode {^{\scriptstyle\prime}} \else $^{\scriptstyle\prime}$\fi}
\newdimen\saa  \newdimen\sbb
\def\parcs{\saa=.07em \sbb=.03em
     \ifmmode \hbox{\rlap{.}}^{\scriptstyle\prime\kern -\sbb\prime}\hbox{\kern -\sa}
     \else \rlap{.}$^{\scriptstyle\prime\kern -\sbb\prime}$\kern -\sa\fi}
\def\parcm{\saa=.08em \sbb=.03em
     \ifmmode \hbox{\rlap{.}\kern\saa}^{\scriptstyle\prime}\hbox{\kern-\sbb}
     \else \rlap{.}\kern\saa$^{\scriptstyle\prime}$\kern-\sbb\fi}

\newcommand{\vlsr}{$v_{\rm LSR}$}
\newcommand{\Rgal}{$R_{\rm gal}$}
\newcommand{\msun}{${\rm M}_{\odot}$}
\newcommand{\SigmaSFR}{$\Sigma_{\rm SFR}$}
\newcommand{\HIGAL}{Hi-GAL}
\usepackage{graphicx}
\usepackage{txfonts}
\begin{document} 

\title{A comparison of the Milky Way's recent star formation revealed by dust thermal emission and high-mass stars}
\titlerunning{Milky Way's star formation rate derived from dust thermal emission and high-masxs stars}
%\subtitle{I. Overviewing the $\kappa$-mechanism}

\author{J.~D.~Soler\inst{1}\thanks{Corresponding author: juandiegosolerp@gmail.com} \and
        E.~Zari\inst{2} \and
        D.~Elia\inst{1} \and
        S.~Molinari\inst{1} \and
        C.~Mininni\inst{1} \and
        E.~Schisano\inst{1} \and
        A.~Traficante\inst{1} \and
        R.~S.~Klessen\inst{3,4} \and
        S.~C.~O.~Glover\inst{3} \and
        P.~Hennebelle\inst{5} \and
        T.~Colman\inst{5} \and 
        N.~Frankel\inst{6} \and
        T.~Wenger\inst{7}
          }

\institute{
1. Istituto di Astrofisica e Planetologia Spaziali (IAPS). INAF. Via Fosso del Cavaliere 100, 00133 Roma, Italy.\\
2. Max-Planck-Institute for Astronomy, K\"{o}nigstuhl 17, 69117 Heidelberg, Germany.\\
3. Universit\"{a}t Heidelberg, Zentrum f\"{u}r Astronomie, Institut f\"{u}r Theoretische Astrophysik, Albert-Ueberle-Str. 2, 69120, Heidelberg, Germany.\\
4. Universit\"{a}t Heidelberg, Interdisziplin\"{a}res Zentrum f\"{u}r Wissenschaftliches Rechnen, 69120 Heidelberg, Germany. \\
5. AIM, CEA, CNRS, Universit\'{e} Paris-Saclay, Universit\'{e} Paris Diderot, Sorbonne Paris Cit\'{e}, 91191 Gif-sur-Yvette, France.\\
6. Canadian Institute for Theoretical Astrophysics, University of Toronto, 60 St. George Street, Toronto, ON M5S 3H8, Canada.\\
7. Department of Astronomy, University of Wisconsin-Madison, 3512 Sterling Hall, Madison, WI 53706-1507, USA.
}
\authorrunning{Soler,\,J.D. et al.}

\date{Submitted 31JUL2023; accepted 11SEP2023}

% \abstract{}{}{}{}{} 
% 5 {} token are mandatory
 
\abstract{
%\LEt{ General notes: A.) I have edited to US English spelling and grammar conventions. B.) A\&A uses the past tense to describe specific methods used in a paper and the present tense to describe general methods as well as findings, including the findings of recent papers (within the past ten or so years). Kindly make any necessary changes. See Sect. 6 of the language guide https://www.aanda.org/for-authors/language-editing/6-verb-tenses. ***}
We present a comparison of the Milky Way's star formation rate (SFR) surface density (\SigmaSFR) obtained with two independent state-of-the-art observational methods.
The first method infers \SigmaSFR\ from observations of the dust thermal emission from interstellar dust grains in far-infrared wavelengths registered in the {\it Herschel} infrared Galactic Plane Survey (\HIGAL). %, as presented in \cite{elia2022}. 
The second method determines \SigmaSFR\ by modeling the current population of O-, B-, and A-type stars in a 6\,kpc\,$\times$\,6\,kpc area around the Sun. %, as presented in \cite{zari2023}.
We find an agreement between the two methods within a factor of two for the mean SFRs and the SFR surface density profiles.
Given the broad differences between the observational techniques and the independent assumptions in the methods for computing the SFRs, this agreement constitutes a significant advance in our understanding of the star formation of our Galaxy and implies that the local SFR has been roughly constant over the past 10\,Myr.
}
\maketitle
%
%----------------------------------------------------

\section{Introduction}\label{sec:intro}

The formation of stars marks the onset of the conversion of nuclear binding energy into radiative and mechanical energy that is then released into the interstellar medium %\LEt{ Verify that your intended meaning has not been changed. ***}
\citep[][]{mckeeANDostriker2007,kennicuttANDevans2012,girichidis2020}.
High-mass stars regulate the composition, structure, and evolution of the interstellar medium  by injecting energy and momentum through supernovae, ionizing photons, and winds \citep[see, for example,][]{weaver1977,krumholz2014}.
Thus, understanding the rate and distribution of star formation is crucial for understanding the workings of the Milky Way and other galaxies \citep[see,][for recent reviews]{klessenANDglover2016,ballesteros-paredes2020,Tacconi2020}.

The Milky Way's star formation rate (SFR) has been estimated using various techniques.
Radio free-free emission and [N{\sc ii}] 205\,$\mu$m emission have been employed to reconstruct the Lyman continuum photon production rate from O stars in the Galactic disk \citep{smith1978}.
Assuming an initial mass function (IMF) of a newly formed generation of stars \citep[][]{salpeter1955,bastian2010,lee2020}, these observations lead to SFR estimates of between 2.0 and 2.4\,M$_{\odot}$\,yr$^{-1}$ \citep{guestenANDmezger1982,bennett1994,mckeeANDwilliams1997,murrayANDrahman2010}.
\cite{robitaille2010} used the census of young stellar objects (YSOs) in the Galactic Legacy Infrared Mid-Plane Survey Extraordinaire (GLIMPSE) survey \citep[][]{benjamin2003} to construct a population synthesis model of the Galaxy, which yields an SFR of between 0.68 and 1.45\,\msun\,yr$^{-1}$.
\cite{chomiukANDpovich2011} normalized various estimates for the Galactic SFR to the same IMF and population synthesis models, which results in a global SFR of around 1.9\,$\pm$\,0.4\,\msun\,yr$^{-1}$.
\cite{licquiaANDnewman2015} revisited this result and, using a hierarchical Bayesian statistical method, obtained 1.65\,$\pm$\,0.19\,\msun\,yr$^{-1}$.
More recently, estimates of the SFR distribution across the Milky Way's disk have become more precise thanks to the advent of high-resolution Galactic plane surveys in radio and far-infrared (FIR) frequencies as well as the unprecedented astrometric observations from the {\it Gaia} space observatory of the European Space Agency (ESA).
%ESA's {\it \LEt{ Consider defining. ***}Gaia} satellite\LEt{ Verify that your intended meaning has not been changed. ***}.

% E22 =====================================
%\subsection{SFR from star-forming clumps}

\citet[][hereafter E22]{elia2022} derived the Milky Way's SFR distribution based on the physical properties of clumps identified in the {\it Herschel}\footnote{{\it Herschel} is an ESA space observatory with science instruments provided by European-led Principal Investigator consortia and with important participation from NASA.} infrared Galactic Plane Survey \citep[\HIGAL;][]{molinari2016}.
\HIGAL\ covered a two-degree-wide strip centered on the midplane of the Galactic disk and registered the emission in 70, 160, 250, 350, and 500\,$\mu$m wavelengths, which are dominated by the thermal emission from interstellar dust grains.
\cite{elia2017,elia2021} identified compact sources in the \HIGAL\ multifrequency maps, which were generically denominated as ``clumps.''

The \HIGAL\ catalog comprises 94\,604 clumps, 35\,186 of which are considered protostellar, that is, at the earliest stage of star formation activity as indicated by their emission at 70\,$\mu$m wavelength \citep{elia2021}.
\cite{mege2021} assigned heliocentric kinematic distances to the clumps by using their morphological matching with the radial velocity channels in carbon monoxide (CO) emission surveys and additional information from three-dimensional dust density reconstructions and neutral atomic hydrogen (H{\sc i}) emission.
Using these distance estimates and the emission registered across the {\it Herschel} wavelength bands, \cite{elia2021} computed the clumps' sizes, masses, bolometric luminosities, and other physical properties.

E22 estimated the SFR for each protostellar clump from an empirical relation between the SFR and the clump mass derived using the method introduced in \cite{veneziani2013,veneziani2017}. 
The initial hypothesis is that the YSO population is related to the SFR in a clump, an assumption that is justified in nearby clouds by the relatively similar ages inferred from the low spread of YSOs in the Hertzsprung-Russell diagram \citep{lada2010}.
The SFR is determined by combining the number of YSOs, their expected final mass, and their time to reach the main sequence.
Models of stellar evolution indicate that a typical timescale estimated for the transition between the YSO phase and the start of the main sequence is roughly between 10$^{5}$ and 10$^{6}$\,yr \citep[see, for example,][]{molinari2008}.

Due to the limited angular resolution, direct YSO counts are limited to the closest sources; thus, the SFR in distant clumps is estimated by applying a phenomenological relation between the YSO counts and the bolometric luminosity obtained in nearby clumps \citep{veneziani2013}.
E22 obtained the relation
\begin{equation}\label{eq:mass2sfr}
{\rm SFR}_{\rm clump} = (5.6 \pm 1.4) \times 10^{-7} (M_{\rm clump}/{\rm M}_{\odot})^{0.74\pm0.03} {\rm M}_{\odot}{\rm yr}^{-1}
\end{equation}
to compute the SFR for 29\,704 protostellar clumps with reliable distance estimates.
The clump SFRs were distributed and summed in a Cartesian grid with 0.5\,kpc\,$\times$\,0.5\,kpc spaxels to calculate the SFR surface density ($\Sigma_{\rm SFR}$) distribution and radial profile within galactocentric radius \Rgal\,$=$\,16\,kpc.

\postref{\cite{wells2022} present a similar derivation of the Milky Way's SFR using dense clumps.
Their analysis is based on modeling the embedded stellar clusters to obtain bolometric luminosities from their parental dust clumps in mid-infrared and FIR frequencies.
The result is a semi-empirical relation between the star formation efficiency and the clump mass, which was applied to obtain the star formation efficiency and SFR from the clumps identified in the 850\,$\mu$m Atacama Pathfinder Experiment (APEX) telescope Large Area Survey of the Galaxy \citep[\mbox{ATLASGAL};][]{schuller2009,urquhart2022}.
The general SFR trends in \cite{wells2022} are not dissimilar to those obtained in E22, that is, the decrease in \SigmaSFR\ for \Rgal\,$>$\,5\,kpc.
However, the Galactic longitude range is limited to $|l|$\,$<$\,60\deg, which restricts their reconstruction of the SFR profile to the inner Galaxy, and their clump selection corresponds to a high-density subset of the \HIGAL\ sample.
Thus, we focus our comparison on the results from E22.}

% Z23 ====================================
%\subsection{SFR from stellar age distribution}

\citet[][hereafter Z23]{zari2023} present a map of the stellar age distribution across a 6\,kpc\,$\times$\,6\,kpc area of the Galactic disk centered on the Sun that was used to reconstruct the Galaxy's recent ($\leq$\,$1$~Gyr) star formation history.
The sample used in Z23 consists of $\sim$ 500\,000 candidate O-, B-, and A-type stars selected in \cite{zari2021} by combining \textit{Gaia} Early Data Release 3 photometry and astrometry \citep{gaia2021A&A...649A...1G} with photometric information from the Two Micron All Sky Survey  \citep[2MASS;][]{skrutskie2006}.
This sample is restricted to sources with \textit{Gaia} G-band magnitudes $<$ 16 and absolute magnitudes in the 2MASS $K_s$ band of less than zero, which roughly corresponds to a late B-type main-sequence star. 
Several color cuts were applied to clean the sample of bright red giant-branch and asymptotic giant-branch stars, as well as objects with unnaturally blue colors.

Z23 modeled the distribution of absolute magnitudes, $M_{K_s}$, that is, $n(M_{K_s}|\vec{x},\,\vec{\alpha})$, taken as a function of position $\vec{x}$ in Cartesian heliocentric coordinates with the $x$-axis oriented in the direction  $l$\,$=$\,0\deg.
They assumed that at each $\vec{x}$ there is an underlying age or birthrate distribution, $b(\tau ~|~ \vec{x}, \vec{\alpha})$, whose temporal dependence is described by a set of parameters, $\vec{\alpha}$, 
\begin{equation}\label{eq:zari23b}
b(\tau~|~\vec{x},\mathbf{\vec\alpha})= \sum_{i} \alpha_{i}(\vec{x})\,\chi_{i}(\tau),
\end{equation}
where the index $i$ runs over five approximately logarithmic age bins:%\LEt{ A\&A discourages bulleted or numbered lists outside of the conclusion, so I have reformatted accordingly. ***}
5 to 10\,Myr, 10 to 30\,Myr, 30 to  100\,Myr, 100 to 300\,Myr, and 300\,Myr to 1\,Gyr.
The indicator function $\chi$\,$=$\,1 if $\tau$ is within the limits of the age bin, and $\chi$\,$=$\,0 otherwise.
For each position, $\vec{x}$, Z23 compared the observed and predicted $M_{K_s}$ distribution and derived the best-fit birthrate parameters, $\vec{\alpha}_{\rm best}$, by optimizing the likelihood of the data.

Z23 hence derived number- or mass-density maps of mono-age stellar populations in five age bins across the 6\,kpc\,$\times$\,6\,kpc area of the Galactic disk surrounding the Sun.
%, as well as the overall radial gradients they show.
To derive a spatially resolved SFR, E23 divided the volume into 256 spaxels of different sizes, each containing roughly 1\,000 stars.
They obtained an estimate of the SFR in that portion of the Galactic disk by summing over the entire area considered and multiplying $b(\tau~|\alpha_{\rm best})$ (units of $\mathrm{yr}^{-1}$) by the mean stellar mass derived from their assumed IMF, $\left<M\right>$\,=\, 0.22\,\msun.

In this paper we compare the independent state-of-the-art methods described above to study the recent star formation history of the Milky Way. 
In Sect.~\ref{sec:comparison} we present the distribution and radial profiles of  the SFR surface density ($\Sigma_{\rm SFR}$) and the comparison between the results of E22 and Z23.
In Sect.~\ref{sec:discussion} we discuss the main differences and similarities between the two estimates. 
Details on the construction of  $\Sigma_{\rm SFR}$ maps and radial profiles are given in Appendix \ref{appendix:method}.

%====================================
\section{Comparison between $\Sigma_{\rm SFR}$ estimates}\label{sec:comparison}

\begin{figure}[ht!]
\centerline{\includegraphics[width=0.499\textwidth,angle=0,origin=c]{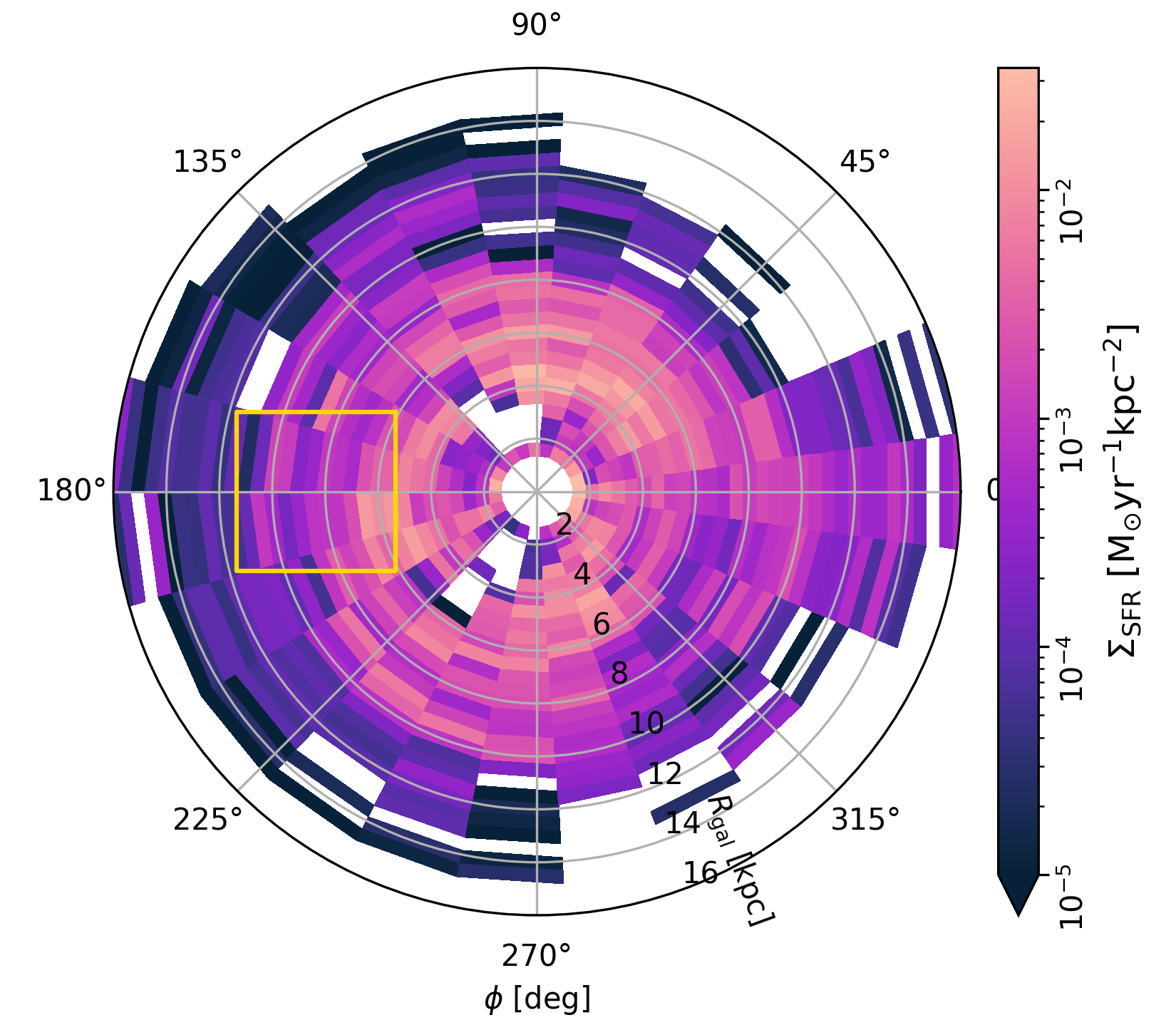}}
\caption{SFR surface density ($\Sigma_{\rm SFR}$) derived from the \HIGAL\ clumps.
The yellow rectangle indicates the position of the 6\,kpc\,$\times$\,6\,kpc area around the Sun studied in Z23.
}
\label{fig:SFRpolargrid}
\end{figure}

We compared the two methods by distributing the E22 protostellar clumps into the spatial grid introduced in Z23, adding their contributions to the SFR to obtain an SFR for each spaxel.
We accounted for the difference in the assumed galactocentric solar orbit radius ($R_{\odot}$) in E22 ($R_{\odot}$\,$=$\,8.35\,kpc) and Z35 ($R_{\odot}$\,$=$\,8.0\,kpc) by displacing the center of the Z23 SFR reconstruction in the local 6\,kpc\,$\times$\,6\,kpc to the Sun position in E22.
The total SFR from E22 and Z23 in this region is $0.13$\,$\pm$\,$0.06$\,M$_{\odot}$\,yr$^{-1}$ and $0.28$\,$\pm$\,$0.05$\,M$_{\odot}$\,yr$^{-1}$, respectively. 

We also compared the \SigmaSFR\ distributions in two ways.
First, we used the \HIGAL\ clumps to compute the \SigmaSFR\ radial profiles over the whole Galactic plane and for the region considered in Z23.
Second, we computed \SigmaSFR\ in each spaxel of the grid introduced in Z23, dividing the SFR by the spaxel area.

% ---------------------------------------------
\subsection{$\Sigma_{\rm SFR}$ radial profiles}

We computed the \HIGAL\ \SigmaSFR\ distribution across the Galactic plane in a polar grid across 1.35\,$<$\,$R_{\rm gal}$\,$<$16.35\,kpc, including the radial bins used by Z23, as shown in Fig. \ref{fig:SFRpolargrid}).
%distributing the \HIGAL\ clumps in a polar grid that coincides with the radial bins used by Z23.
We calculated \SigmaSFR\ by adding the SFRs from the \HIGAL\ clumps in each spaxel and diving by the area of the spaxel.
We employed the clump positions in Galactic coordinates and the heliocentric distances reported in E22's catalog to compute the galactocentric radius (\Rgal) using the astronomical coordinate transformation tools in {\tt astropy} \citep{astropy:2022}.
Further details on the population of the grid are provided in Appendix~\ref{appendix:method}.

The \SigmaSFR\ distribution, presented in Fig.~\ref{fig:SFRpolargrid}, illustrates the significant decrease in \SigmaSFR\ with increasing \Rgal\ already identified in E22.
It also shows a decrease in \SigmaSFR\ of at least one order of magnitude from the lowest to the largest \Rgal\ within the area considered in Z23.
The largest \SigmaSFR\ values within that area considered in Z23 are concentrated toward the lower-right corner of the region within the yellow rectangle in Fig. \ref{fig:SFRpolargrid}, which corresponds to the fourth Galactic quadrant for an observer located at the Sun's position.

\begin{figure}[ht!]
\centerline{\includegraphics[width=0.49\textwidth,angle=0,origin=c]{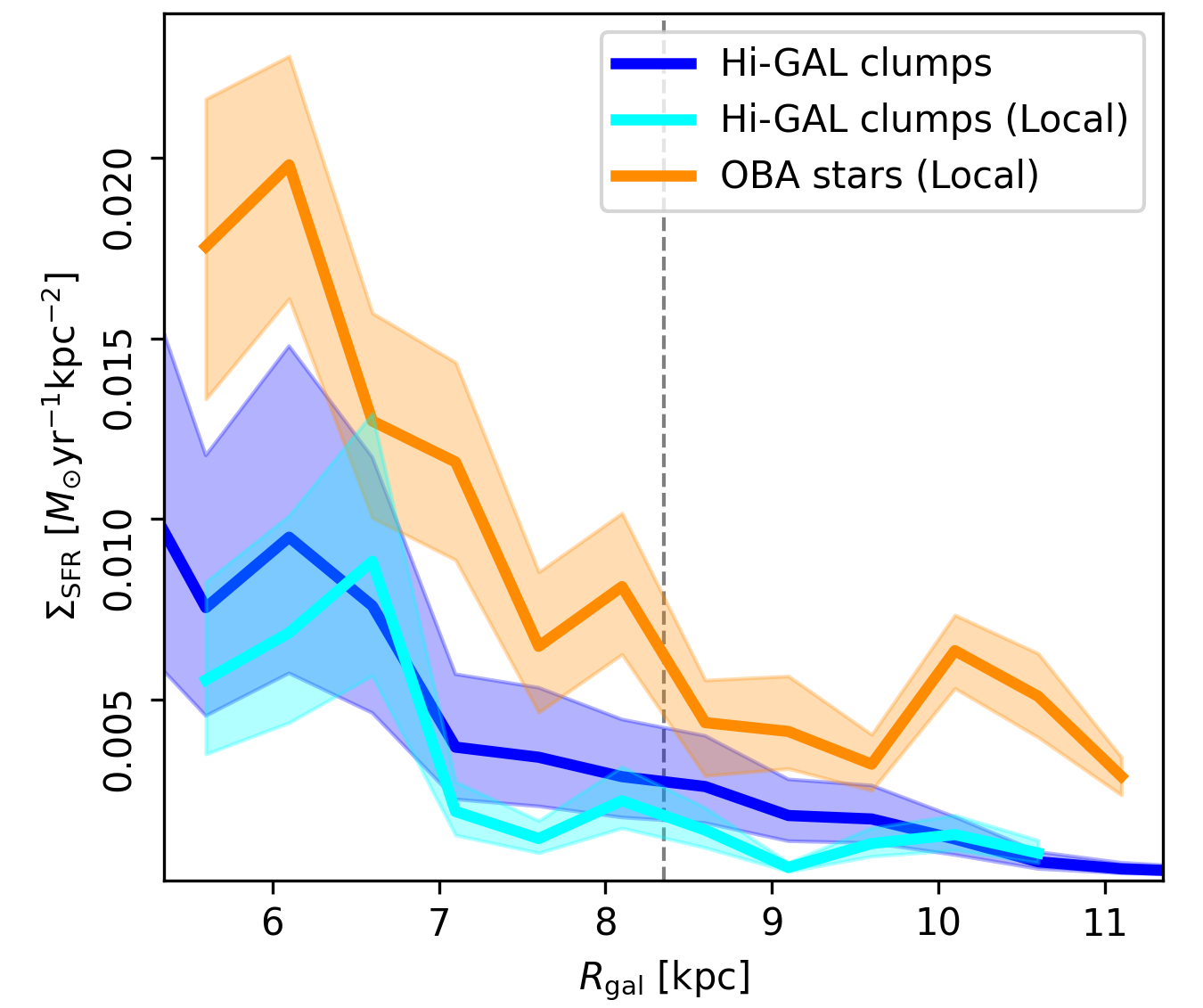}}
\centerline{\includegraphics[width=0.49\textwidth,angle=0,origin=c]{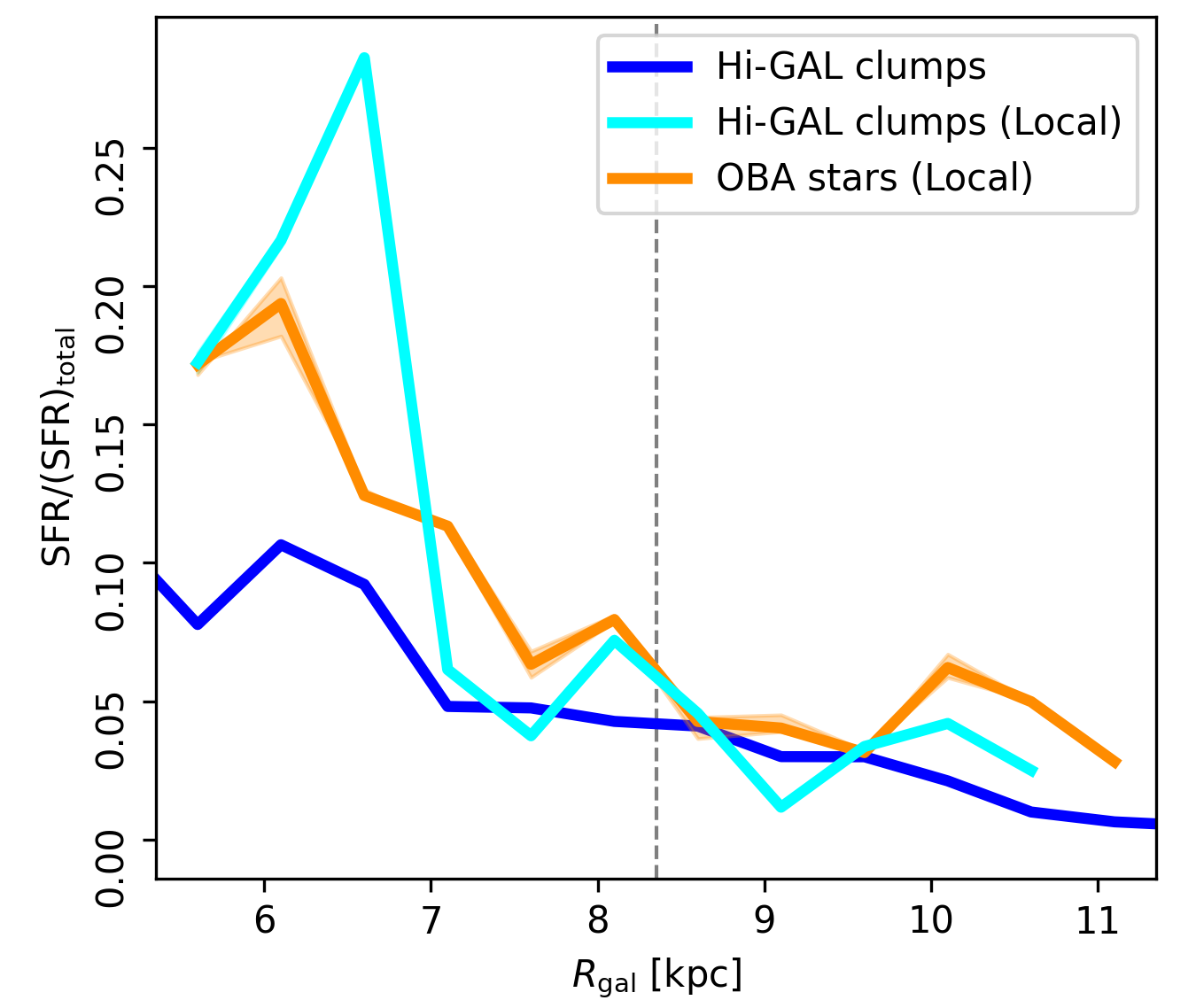}}
\caption{Profiles of the star formation surface density ($\Sigma_{\rm SFR}$; top) and normalized star formation (SFR/SFR$_{\rm total}$, where SFR$_{\rm total}$\,$\equiv$\,$\Sigma^{R_{\rm max}}_{R_{\rm min}}{\rm SFR}$; bottom) obtained from modeling of the populations of high-mass stars in the age interval $5$\,$<$\,$\tau$\,$<$\,10\,Myr \citep[orange;][]{zari2023} and the FIR emission from the \HIGAL\ clumps \citep[blue and cyan;][]{elia2022}.
The orange and cyan curves, labeled ``local,'' correspond to the profiles derived from an area of 6\,kpc\,$\times$\,6\,kpc around the Sun.
The blue lines correspond to the profile from the full azimuthal range. 
The vertical segmented line indicates the assumed position of the Sun.
The shaded areas indicate the 16th and 84th percentiles in the \cite{zari2023} estimates and the upper and lower limits given by Eq.~\ref{eq:mass2sfr} in the E22 estimates.
}
\label{fig:SFRprofiles}
\end{figure}

We calculated the Galactic \SigmaSFR\ radial profiles by binning and adding the SFRs from the \HIGAL\ clumps in the radial grid used to construct Fig.~\ref{fig:SFRpolargrid} and dividing by the area of each annulus.
For the local 6\,kpc\,$\times$\,6\,kpc considered in Z23 (yellow box in Fig.~\ref{fig:SFRpolargrid}), we added the SFRs from the \HIGAL\ clumps within the intersection between each annulus and the square patch and divided by the resulting area. 
The top panel of Fig.~\ref{fig:SFRprofiles} shows the radial profiles obtained from the \HIGAL\ clumps and the results of Z23 for the age interval $5$\,$<$\,$\tau$\,$<$\,10\,Myr.
We chose this age range since it is the closest to the timescales sampled by the FIR emission. 
Furthermore, the profiles obtained in other age bins are relatively flat (see Z23).

The global profile derived from the \HIGAL\ clumps is around a factor of two below that obtained in Z23, but it shows a similar dependence on \Rgal.
This profile is particularly dominated by the significantly large \SigmaSFR\ shown by both methods for \Rgal\,$<$\,8\,kpc.
The radial profiles corresponding to the local 6\,kpc\,$\times$\,6\,kpc region show similar features at different galactocentric radii, as highlighted in the normalized SFR profile shown in the bottom panel of Fig.~\ref{fig:SFRprofiles}.
For \Rgal\,$\gtrsim$\,7\,kpc, both tracers show significantly lower $\Sigma_{\rm SFR}$ than at smaller $R_{\rm gal}$, a bump in $\Sigma_{\rm SFR}$ around $R_{\rm gal}$\,$\approx$\,7.8\,kpc, and a second bump at $R_{\rm gal}$\,$\gtrsim$\,7.8\,kpc.
For $R_{\rm gal}$\,$\lesssim$\,7\,kpc, both profiles show an increase in $\Sigma_{\rm SFR}$  and a decrease close to $R_{\rm gal}$\,$\approx$\,5\,kpc.
We further explored the structure in the $\Sigma_{\rm SFR}$ profiles by considering the distribution within the area studied in Z23.

% ----------------------------------------------------------------------
\subsection{Local $\Sigma_{\rm SFR}$ distribution}

\begin{figure*}[ht!]
\centerline{\includegraphics[width=0.99\textwidth,angle=0,origin=c]{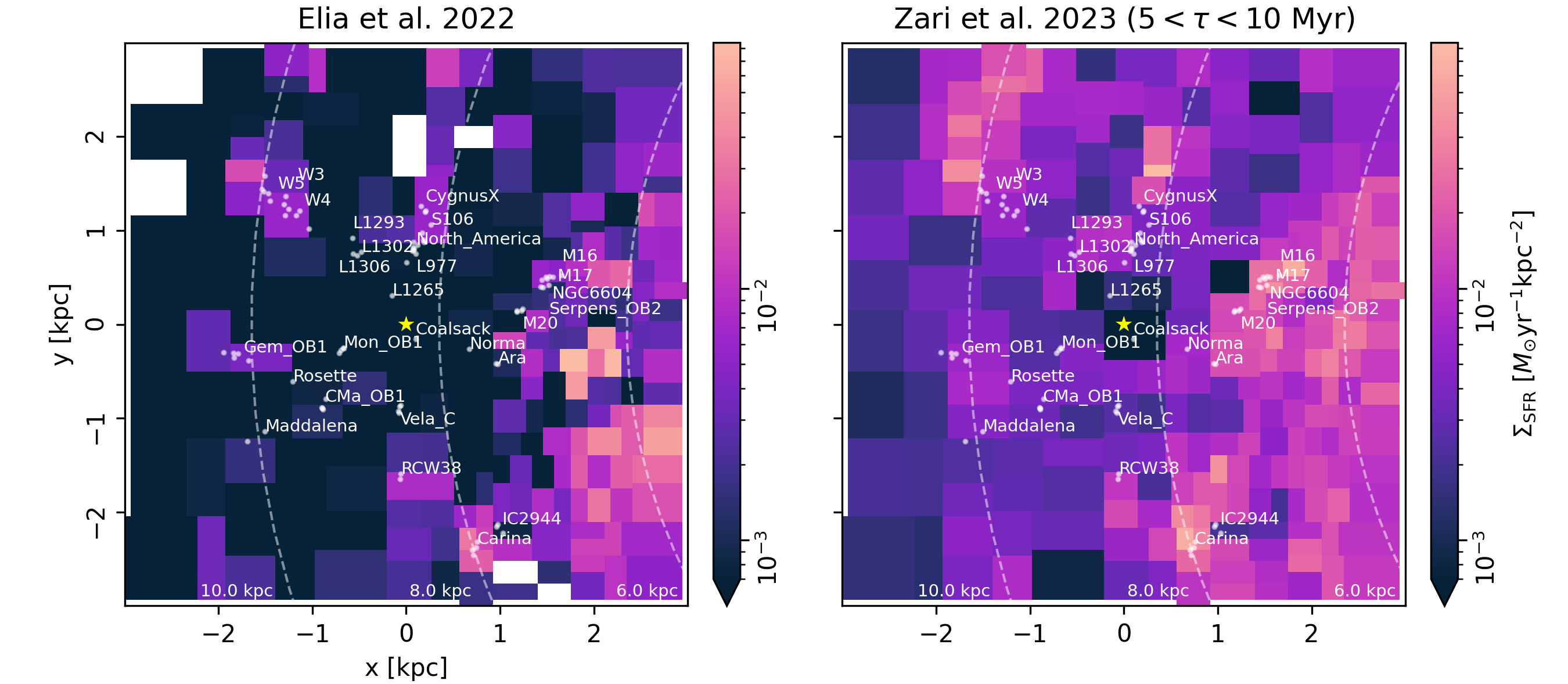}}
\caption{SFR surface density ($\Sigma_{\rm SFR}$) obtained from the \HIGAL\ clumps (left) and modeling of the high-mass stellar population (right) in a 6\,kpc\,$\times$\,6\,kpc area centered on the Sun (indicated with the yellow star).
The labels give the positions of the sources in the ``Handbook of Star-Forming Regions'' \citep{reipurth2008a,reipurth2008b} with the distances estimated in \cite{zucker2020}.
The dashed lines correspond to galactocentric circles with the indicated radii. 
}
\label{fig:SFRmaps}
\end{figure*}

We also computed the \SigmaSFR\ distribution from the \HIGAL\ clumps using the same spatial grid as in Z23.
As in the polar grid, we added the SFRs for the clumps in each spaxel and divided them by the spaxel area to obtain \SigmaSFR.
We present the results of the two \SigmaSFR\ estimates in Fig.~\ref{fig:SFRmaps}.

Figure~\ref{fig:SFRmaps} shows variations of more than an order of magnitude in \SigmaSFR\ as a function of position in the Galaxy.
Such variations are also observed in nearby spiral galaxies \citep{kreckel2018,sun2023}.
They imply that the instantaneous SFR varies dramatically throughout the Galactic disk.

The main similarities between the \SigmaSFR\ distributions are found toward well-known high-mass star-forming regions (counterclockwise and starting from the top of Fig.~\ref{fig:SFRmaps}): Cygnus \citep{reipurth2008cygnus}, the W3/W4/W5 region \citep{megeath2008}, RCW38 \citep{wolk2008}, and the Carina Nebula \citep{smith2008}.
The agreement is also good in the relatively high $\Sigma_{\rm SFR}$ toward the M16, M17, M20, and Gem OB1 regions.
The latter agreement is particularly notable since the uncertainty in the kinematic distances is particularly acute toward the Galactic center and anticenter.
%\LEt{ Verify that your intended meaning has not been changed. ***}

\begin{figure}[ht!]
\centerline{\includegraphics[width=0.49\textwidth,angle=0,origin=c]{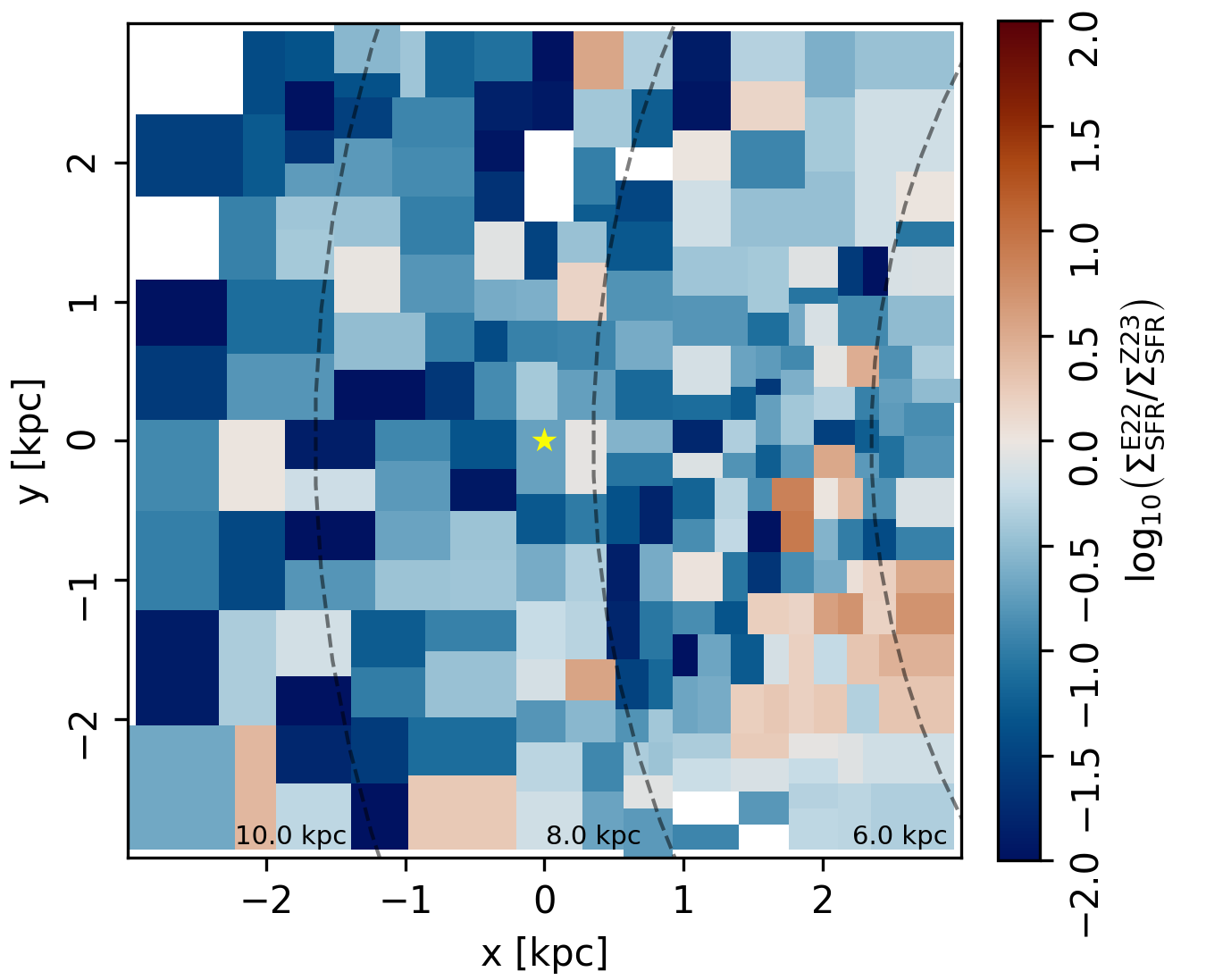}}
\caption{Ratio between the \SigmaSFR\ estimates derived in E22 and Z23.}
\label{fig:SFRcomparison}
\end{figure}

Figure~\ref{fig:SFRcomparison} shows the ratio of the two star formation surface density estimates $\Sigma^{\rm E22}_{\rm SFR}/\Sigma^{\rm Z23}_{\rm SFR}$ across the local patch presented in Fig.~\ref{fig:SFRmaps}.
The $\Sigma^{\rm E22}_{\rm SFR}/\Sigma^{\rm Z23}_{\rm SFR}$ distribution indicates that the \SigmaSFR\ estimates from Z23 are larger in most of the spaxels.
There are, however, some regions where the \SigmaSFR\ from E22 are the largest, for example, toward the lower-left quadrant of the local patch (fourth Galactic quadrant).

The values of $\Sigma^{\rm E22}_{\rm SFR}/\Sigma^{\rm Z23}_{\rm SFR}$ shown in Fig.~\ref{fig:SFRcomparison} appear to be consistently lower toward higher \Rgal.
Yet, there is no global trend in the distribution of the ratio, and regions at roughly the same \Rgal\ show either $\Sigma^{\rm E22}_{\rm SFR}/\Sigma^{\rm Z23}_{\rm SFR}$\,$>$\,$1$ or $<$\,$1$, particularly for \Rgal\,$<$\,8\,kpc.
%\LEt{ Verify that your intended meaning has not been changed. ***}
In the following section we discuss whether physical or systematic effects cause these differences.

% ==============================================
\section{Discussion}
\label{sec:discussion}

We have presented \SigmaSFR\ as derived from the \HIGAL\ clumps and the distribution of the high-mass stars, focusing on the end product of the two observational methods.
In this section we discuss whether the differences between the two estimates are explained by a systematic effect introduced by the underlying assumptions in either method or whether they reveal further physical insights into the dynamics of star formation in the Galactic plane.

% ----------------------------------------------------
\subsection{Why the $\Sigma_{\rm SFR}$ distributions are similar}

%\LEt{ A\&A encourages authors to avoid directly addressing the reader in the main text, e.g., by using the imperative or posing direct questions. ***}
The agreement between the \SigmaSFR\ profiles reported in Fig.~\ref{fig:SFRprofiles} is noteworthy, given the differences in the observed quantities and the underlying assumptions involved in the two methods.
However, this similarity is not entirely unexpected.

% Kennicutt-Schmidt argument
The molecular gas mass surface density ($\Sigma_{{\rm H}_{2}}$) traced by CO shows a sharp increase toward the inner Galaxy, reaching a peak value around 4\,$<$\,$R_{\rm gal}$\,$<$\,5\,kpc \citep[see, for example,][]{heyerANDdame2015,miville-deschenes2017,riener2020}.
The star formation surface density is also expected to increase toward the inner Galaxy, directly following the strong correlation between $\Sigma_{\rm SFR}$ and $\Sigma_{{\rm H}_{2}}$ observed toward nearby disk galaxies \citep[see, for example,][]{kennicutt1989,schruba2011,leroy2013}.
Hence, it is expected that \SigmaSFR\ increases at small \Rgal. 

% Timescales are similar
Z23 concluded that the Milky Way's SFR increased during the past 10\,Myr. 
Thus, it is foreseeable that the more recent SFR traced by the \HIGAL\ clumps follows a similar distribution.
If we assume that the difference between the two \SigmaSFR\ profiles in Fig.~\ref{fig:SFRprofiles} has a physical origin, it would mean that the Milky Way SFR peaked between 5 to 10\,Myr ago and has since slightly decreased. 
Yet, there are many systematic effects that may reconcile the two estimates without implying a change in the star formation history, as we discuss below.  

% -----------------------------------------
\subsection{Why the $\Sigma_{\rm SFR}$ \postref{values} are different}

The underlying assumptions in the estimation of \SigmaSFR\ are likely explanations for the factor of around two separating the total SFR and the \SigmaSFR\ profiles in Fig.~\ref{fig:SFRprofiles}.
On the \HIGAL\ clump SFR estimate, these assumptions include the use of kinematic distances, the selection of the dust properties to reconstruct masses from FIR emission, and the phenomenological model used to map masses into SFR.
On the high-mass stellar population modeling, these assumptions include the adopted completeness of the sample and the choice of IMF.

% Kinematic distances
The kinematic distances of the \HIGAL\ clumps are derived by matching their emission to a particular component of CO emission \citep{mege2021}.
If the clump is on a circular orbit following a Galactic rotation model, the velocity with respect to the local standard of rest (\vlsr) uniquely identifies the clump's \Rgal\ \citep[see, for example,][]{wegner2018}.
This means that clumps close to the Galactic center or anticenter, where the radial component of the rotational motion is minimum, will have large uncertainties in their kinematic distances. 
Moreover, non-radial motions, such as those arising from supernova explosions or spiral arm shocks, will affect the location of clumps based on kinematic distances \citep[see, for example,][]{burton1971,peek2022,hunter2023}.
The \SigmaSFR\ estimates from Z23 do not depend on kinematic distances but from {\it Gaia} parallaxes.
Thus, because the two profiles in Fig.~\ref{fig:SFRprofiles} are offset in \SigmaSFR\ rather than displaced in \Rgal, it is likely that the kinematic distances are not critical for the global reconstruction of the SFR.

% Masses from dust 
The SFR from the \HIGAL\ clumps depends on the mass estimates, which were obtained using the source size, assumed distance, emission optical depth ($\tau$), and dust opacity ($\kappa$) \citep[Eq.~3 in][]{elia2017}.
The value of $\tau$ is usually parametrized as a power law with spectral index $\beta$.
Following the {\it Herschel} observations in nearby clouds regions \citep{sadavoy2013,konyves2015}, E22 adopted the reference value $\beta$\,$=$\,2.
However, variations of $\beta$ between 1.8 and 2.6 have been inferred from {\it Herschel} observations toward the Galactic plane \citep{paradis2010}.
An increase in the assumed values of $\beta$ would push toward higher masses and higher SFRs, consequently closing the gap between the results of E22 and Z23.
Yet, the selection of a single high $\beta$ for all the clumps is not justified by the observational evidence \citep[see, for example,][]{planck2013-p06b}.

E22 set $\kappa$ to a reference value at 300\,$\mu$m, $\kappa_{300}$\,$=$\,$0.1$\,cm$^{2}$\,g$^{-1}$, which includes the underlying assumption of a dust-to-gas ratio $\gamma$\,$=$\,100 \citep{beckwith1990}.
Dust models indicate $\kappa$ values that span from $\kappa_{300}$\,$=$\,0.13\,cm$^{2}$\,g$^{-1}$ in \cite{ossenkopf1994} to one of magnitude lower in \cite{draineANDlee2007}.
Assuming a lower value of $\kappa$ within the current observational limits is enough to reconcile the offset in Fig.~\ref{fig:SFRprofiles}. 
Fluctuations in $\kappa$ are expected from the variety of physical conditions among the clumps, but it is unlikely that they conspire to make $\kappa$ systematically lower than the reference value used in E22 and solely explain the offset in the \SigmaSFR\ profiles.

% Phenomenological model
The phenomenological relation in Eq.~\eqref{eq:mass2sfr} was obtained using plausible but coarse assumptions about the evolution between the protostellar phase and the start of the main sequence \citep{molinari2008}.
Assuming longer (shorter) timescales leads to an underestimation (overestimation) of the SFR.
A shorter timescale would reconcile the \SigmaSFR\ profiles, although there is no observational evidence that would justify this choice.

% SFR from stars higher than SFR from clumps
Figure~\ref{fig:SFRmaps} shows that the \SigmaSFR\ as derived from the \HIGAL\ clumps is lower than \SigmaSFR\ as derived from high-mass stars.
The difference is more pronounced outside of the known star-forming regions.
The spaxels showing good agreement \postref{most likely} correspond to \postref{regions} where the massive stars and YSOs coexist, while the spaxels with the largest differences correspond to regions with less dust and more stars.

% Masses underestimated in HIGAL
The minimum mass recovered with the \HIGAL\ survey is limited by the sensitivity of the {\it Herschel} instruments.
This means that in regions of the sky with low FIR emission, the masses and SFRs derived from the \HIGAL\ observations are underestimated.
This may be the case toward the highest $R_{\rm gal}$ regions and the lower-left quadrant in Fig.~\ref{fig:SFRmaps} and the area between the W3/W4/W5 complex and Cygnus X.
Additionally, the catalog used in E22 includes only ``cold'' clumps, that is, objects with reliable spectral energy density between 160 and 500\,$\mu$m, from which their mass is derived.
Multiple 70\,$\mu$m-, 70 and 160\,$\mu$m-, and 70, 160, and 250\,$\mu$m-only sources are excluded from the SFR tally.
Furthermore, the 70 $\mu$m emission employed to determine whether or not a clump is protostellar, and thus its inclusion in the SFR calculation, is an ``on-off switch''; clumps with 70 $\mu$m emission close or below the detection limit are also excluded.
Finally, clumps without reliable heliocentric distances are also not included.
%The SFR contribution from the latter group alone could account for roughly an additional 20\% to the global Galactic SFR.
\postref{These omitted sources could account for an additional 20\% to the global Galactic SFR, as estimated using random distance realizations \cite[see Sect. 2.2.1 in][]{elia2022}.}

% Stellar dispersion
The discrepancy between the two \SigmaSFR\ estimates can also be caused by the fact that some O-, B-, and A-type stars might have moved from their birthplaces and populated the regions where the \HIGAL\ emission is low, a hypothesis that is supported by the current observational evidence.
The velocity dispersion of YSOs in the Orion high-mass star formation region is around 5\,km\,s$^{-1}$, which corresponds to an average displacement of roughly 50\,pc in 10\,Myr \citep{grossschedl2021}.
Furthermore, stars in the Galactic disk deviate from their birth orbits and experience radial migration and vertical heating.

Although based on older stars, the model of the secular evolution of the Milky Way disk presented in \cite{frankel2020} indicates that the expected radial migration close to the solar circle in 10\,Myr is around 100\,pc. 
This is further supported by recent hydrodynamical simulations presented in \cite{Fujimoto2023}, where the numerical experiment set up to investigate the effect of gravitational interactions between giant molecular clouds (GMCs) and young stars in a Galactic disk analog indicates that GMCs efficiently scatter newborn stars in the first several hundred Myr after the stellar birth (see their Fig.~7).
Moreover, observations of nearby spiral galaxies suggest that GMCs exist for less than roughly 5\,Myr after the onset of massive star formation \citep[see, for example,][]{kim2022}.
Together, these pieces of evidence suggest that the 5 to 10\,Myr old OB stars should be located in different places than the ongoing star formation and thus partly explain why there is not a tighter agreement between the \SigmaSFR\ distribution presented in Fig.~\ref{fig:SFRmaps}.

% IMF
Finally, the choice of the IMF employed to model the stellar birth rate in Z23 can globally reconcile the profiles in Fig.~\ref{fig:SFRprofiles}.
Since their sample is predominantly composed of massive stars, of which most are in unresolved binaries, Z23 used a four-component broken power law that accounts for unresolved binaries from \citet{kroupa2002}.
%$M^{-\Gamma}$ with $\Gamma$\,$=$\,2.3 for masses between 1 and 120\,\msun.
However, a two-component broken power law with $\Gamma$\,$=$\,$1.3$ for masses between 0.1 and 0.5\,$M_{\odot}$ and $\Gamma$\,$=$\,$2.3$ for masses between 0.5 and 100\,$M_{\odot}$ \citep{Kroupa2003}, leads to much better agreement with the E22 \SigmaSFR\ profile, as we show in Appendix~\ref{appendix:IMF}.
This agreement does not justify the selection of a two-component IMF, but illustrates the quantitative agreement between the \SigmaSFR\ profiles within the range of assumptions of both observational methods.

%-----------------------------------------------------------------
\section{Conclusions}

% General result
We have presented a comparison of the Milky Way's \SigmaSFR\ spatial distribution and radial profile inferred from two independent observational methods: the modeling of high-mass stellar populations and the FIR dust thermal emission from the clumps of the \HIGAL\ survey. 
The two estimates show good agreement in the radial \SigmaSFR\ profile, within a factor of two, and considerable agreement in the SFR distribution within the 6\,kpc\,$\times$\,6\,kpc area around the Sun.
The factor of two separating the SFR estimates can be easily accounted for by the assumptions underlying the two methods.
Together, these results indicate that the timescales of star formation sampled by each method are similar and suggest that the local SFR in the past 10\,Myr has been approximately constant.
%in the area within 3\,kpc around the Sun.

The agreement between the two independent SFR estimates presented here supports the robustness of both SFR reconstructions. %\LEt{ Verify that your intended meaning has not been changed. ***}
The \HIGAL\ results suggest that the conclusions drawn from the local populations of high-mass stars can be extended to the rest of the Galaxy.
The outcome of the high-mass stellar population modeling indicates that the analysis of the dust thermal emission is not critically affected by the assumption of kinematic distances and dust properties.
Given the difficulties derived from our position within the Galaxy, the agreement between these two estimates is comparable to characters in the parable of blind people encountering an elephant\footnote{{\it ``It was six men of Indostan\\
To learning much inclined,\\
Who went to see the Elephant\\
(Though all of them were blind),\\
That each by observation\\
Might satisfy his mind...''} \citep{saxe1868poems}} 
and agreeing with each other on their observations. %\LEt{ Verify that your intended meaning has not been changed. ***}
In this case, the observations constitute a crucial hint about the workings of the Milky Way and a significant point of comparison with other nearby spiral galaxies.
    
\begin{acknowledgements}
JDS, SM, CM, RSK, PH, SCOG, and TC acknowledge funding by the European Research Council via the ERC Synergy Grant ``ECOGAL -- Understanding our Galactic ecosystem: From the disk of the Milky Way to the formation sites of stars and planets'' (project ID 855130).
RSK and SCOG furthermore acknowledge funding from the German Excellence Strategy via the Heidelberg Cluster of Excellence (EXC 2181 - 390900948) ``STRUCTURES'' and from the German Ministry for Economic Affairs and Climate Action in project ``MAINN'' (funding ID 50OO2206). 
The team in Heidelberg also thanks for the computing resources provided by {\em The L\"{a}nd} and DFG through grant INST 35/1134-1 FUGG and for data storage at SDS@hd through grant INST 35/1314-1 FUGG.
This project was \postref{partly developed} at the Heidelberg-Harvard Star Formation workshop held in Heidelberg, Germany, in December 2022 and at the Lorentz Center workshop ``Mapping the Milky Way'', held in Leiden, Netherlands, in February 2023.
The authors acknowledge Interstellar Institute’s program ``II6'' and the Paris-Saclay University’s Institut Pascal for hosting discussions that nourished the development of the ideas behind this work.
\postref{We thank the anonymous referee for the insightful comments that improved the quality of this paper.}
\postref{We also thank James Urquhart for his guidance through the ATLASGAL papers.}
JDS thanks the following people who helped with their encouragement and conversation: Jo\~{a}o Alves, Henrik Beuther, Neal~J.~Evans~II, Alyssa Goodman, Josefa Gro{\ss}schedl, Marc-Antoine Miville-Desch\^{e}nes, Cameren Swiggum, and Catherine Zucker.
\end{acknowledgements}

\bibliographystyle{aa}
%\bibliography{sfrProfile.bib}
\bibliography{sfrProfile.bbl}

\begin{thebibliography}{70}
\expandafter\ifx\csname natexlab\endcsname\relax\def\natexlab#1{#1}\fi

\bibitem[{{Astropy Collaboration} {et~al.}(2022){Astropy Collaboration},
  {Price-Whelan}, {Lim}, {Earl}, {Starkman}, {Bradley}, {Shupe}, {Patil},
  {Corrales}, {Brasseur}, {N{"o}the}, {Donath}, {Tollerud}, {Morris},
  {Ginsburg}, {Vaher}, {Weaver}, {Tocknell}, {Jamieson}, {van Kerkwijk},
  {Robitaille}, {Merry}, {Bachetti}, {G{"u}nther}, {Aldcroft},
  {Alvarado-Montes}, {Archibald}, {B{'o}di}, {Bapat}, {Barentsen}, {Baz{'a}n},
  {Biswas}, {Boquien}, {Burke}, {Cara}, {Cara}, {Conroy}, {Conseil}, {Craig},
  {Cross}, {Cruz}, {D'Eugenio}, {Dencheva}, {Devillepoix}, {Dietrich},
  {Eigenbrot}, {Erben}, {Ferreira}, {Foreman-Mackey}, {Fox}, {Freij}, {Garg},
  {Geda}, {Glattly}, {Gondhalekar}, {Gordon}, {Grant}, {Greenfield}, {Groener},
  {Guest}, {Gurovich}, {Handberg}, {Hart}, {Hatfield-Dodds}, {Homeier},
  {Hosseinzadeh}, {Jenness}, {Jones}, {Joseph}, {Kalmbach}, {Karamehmetoglu},
  {Ka{l}uszy{'n}ski}, {Kelley}, {Kern}, {Kerzendorf}, {Koch}, {Kulumani},
  {Lee}, {Ly}, {Ma}, {MacBride}, {Maljaars}, {Muna}, {Murphy}, {Norman},
  {O'Steen}, {Oman}, {Pacifici}, {Pascual}, {Pascual-Granado}, {Patil},
  {Perren}, {Pickering}, {Rastogi}, {Roulston}, {Ryan}, {Rykoff}, {Sabater},
  {Sakurikar}, {Salgado}, {Sanghi}, {Saunders}, {Savchenko}, {Schwardt},
  {Seifert-Eckert}, {Shih}, {Jain}, {Shukla}, {Sick}, {Simpson},
  {Singanamalla}, {Singer}, {Singhal}, {Sinha}, {Sip{H{o}}cz}, {Spitler},
  {Stansby}, {Streicher}, {{{S}}umak}, {Swinbank}, {Taranu}, {Tewary},
  {Tremblay}, {Val-Borro}, {Van Kooten}, {Vasovi{'c}}, {Verma}, {de Miranda
  Cardoso}, {Williams}, {Wilson}, {Winkel}, {Wood-Vasey}, {Xue}, {Yoachim},
  {Zhang}, {Zonca}, \& {Astropy Project Contributors}}]{astropy:2022}
{Astropy Collaboration}, {Price-Whelan}, A.~M., {Lim}, P.~L., {et~al.} 2022,
  apj, 935, 167

\bibitem[{{Ballesteros-Paredes} {et~al.}(2020){Ballesteros-Paredes},
  {Andr{\'e}}, {Hennebelle}, {Klessen}, {Kruijssen}, {Chevance}, {Nakamura},
  {Adamo}, \& {V{\'a}zquez-Semadeni}}]{ballesteros-paredes2020}
{Ballesteros-Paredes}, J., {Andr{\'e}}, P., {Hennebelle}, P., {et~al.} 2020,
  \ssr, 216, 76

\bibitem[{{Bastian} {et~al.}(2010){Bastian}, {Covey}, \& {Meyer}}]{bastian2010}
{Bastian}, N., {Covey}, K.~R., \& {Meyer}, M.~R. 2010, \araa, 48, 339

\bibitem[{{Beckwith} {et~al.}(1990){Beckwith}, {Sargent}, {Chini}, \&
  {Guesten}}]{beckwith1990}
{Beckwith}, S. V.~W., {Sargent}, A.~I., {Chini}, R.~S., \& {Guesten}, R. 1990,
  \aj, 99, 924

\bibitem[{{Benjamin} {et~al.}(2003){Benjamin}, {Churchwell}, {Babler}, {Bania},
  {Clemens}, {Cohen}, {Dickey}, {Indebetouw}, {Jackson}, {Kobulnicky},
  {Lazarian}, {Marston}, {Mathis}, {Meade}, {Seager}, {Stolovy}, {Watson},
  {Whitney}, {Wolff}, \& {Wolfire}}]{benjamin2003}
{Benjamin}, R.~A., {Churchwell}, E., {Babler}, B.~L., {et~al.} 2003, \pasp,
  115, 953

\bibitem[{{Bennett} {et~al.}(1994){Bennett}, {Fixsen}, {Hinshaw}, {Mather},
  {Moseley}, {Wright}, {Eplee}, {Gales}, {Hewagama}, {Isaacman}, {Shafer}, \&
  {Turpie}}]{bennett1994}
{Bennett}, C.~L., {Fixsen}, D.~J., {Hinshaw}, G., {et~al.} 1994, \apj, 434, 587

\bibitem[{{Burton}(1971)}]{burton1971}
{Burton}, W.~B. 1971, \aap, 10, 76

\bibitem[{{Chomiuk} \& {Povich}(2011)}]{chomiukANDpovich2011}
{Chomiuk}, L. \& {Povich}, M.~S. 2011, \aj, 142, 197

\bibitem[{{Draine} \& {Li}(2007)}]{draineANDlee2007}
{Draine}, B.~T. \& {Li}, A. 2007, \apj, 657, 810

\bibitem[{{Elia} {et~al.}(2021){Elia}, {Merello}, {Molinari}, {Schisano},
  {Zavagno}, {Russeil}, {M{\`e}ge}, {Martin}, {Olmi}, {Pestalozzi}, {Plume},
  {Ragan}, {Benedettini}, {Eden}, {Moore}, {Noriega-Crespo}, {Paladini},
  {Palmeirim}, {Pezzuto}, {Pilbratt}, {Rygl}, {Schilke}, {Strafella}, {Tan},
  {Traficante}, {Baldeschi}, {Bally}, {di Giorgio}, {Fiorellino}, {Liu},
  {Piazzo}, \& {Polychroni}}]{elia2021}
{Elia}, D., {Merello}, M., {Molinari}, S., {et~al.} 2021, \mnras, 504, 2742

\bibitem[{{Elia} {et~al.}(2017){Elia}, {Molinari}, {Schisano}, {Pestalozzi},
  {Pezzuto}, {Merello}, {Noriega-Crespo}, {Moore}, {Russeil}, {Mottram},
  {Paladini}, {Strafella}, {Benedettini}, {Bernard}, {Di Giorgio}, {Eden},
  {Fukui}, {Plume}, {Bally}, {Martin}, {Ragan}, {Jaffa}, {Motte}, {Olmi},
  {Schneider}, {Testi}, {Wyrowski}, {Zavagno}, {Calzoletti}, {Faustini},
  {Natoli}, {Palmeirim}, {Piacentini}, {Piazzo}, {Pilbratt}, {Polychroni},
  {Baldeschi}, {Beltr{\'a}n}, {Billot}, {Cambr{\'e}sy}, {Cesaroni},
  {Garc{\'\i}a-Lario}, {Hoare}, {Huang}, {Joncas}, {Liu}, {Maiolo}, {Marsh},
  {Maruccia}, {M{\`e}ge}, {Peretto}, {Rygl}, {Schilke}, {Thompson},
  {Traficante}, {Umana}, {Veneziani}, {Ward-Thompson}, {Whitworth}, {Arab},
  {Bandieramonte}, {Becciani}, {Brescia}, {Buemi}, {Bufano}, {Butora},
  {Cavuoti}, {Costa}, {Fiorellino}, {Hajnal}, {Hayakawa}, {Kacsuk}, {Leto}, {Li
  Causi}, {Marchili}, {Martinavarro-Armengol}, {Mercurio}, {Molinaro},
  {Riccio}, {Sano}, {Sciacca}, {Tachihara}, {Torii}, {Trigilio}, {Vitello}, \&
  {Yamamoto}}]{elia2017}
{Elia}, D., {Molinari}, S., {Schisano}, E., {et~al.} 2017, \mnras, 471, 100

\bibitem[{{Elia} {et~al.}(2022){Elia}, {Molinari}, {Schisano}, {Soler},
  {Merello}, {Russeil}, {Veneziani}, {Zavagno}, {Noriega-Crespo}, {Olmi},
  {Benedettini}, {Hennebelle}, {Klessen}, {Leurini}, {Paladini}, {Pezzuto},
  {Traficante}, {Eden}, {Martin}, {Sormani}, {Coletta}, {Colman}, {Plume},
  {Maruccia}, {Mininni}, \& {Liu}}]{elia2022}
{Elia}, D., {Molinari}, S., {Schisano}, E., {et~al.} 2022, \apj, 941, 162

\bibitem[{{Frankel} {et~al.}(2020){Frankel}, {Sanders}, {Ting}, \&
  {Rix}}]{frankel2020}
{Frankel}, N., {Sanders}, J., {Ting}, Y.-S., \& {Rix}, H.-W. 2020, \apj, 896,
  15

\bibitem[{{Fujimoto} {et~al.}(2023){Fujimoto}, {Inutsuka}, \&
  {Baba}}]{Fujimoto2023}
{Fujimoto}, Y., {Inutsuka}, S.-i., \& {Baba}, J. 2023, \mnras, 523, 3049

\bibitem[{{Gaia Collaboration} {et~al.}(2021){Gaia Collaboration}, {Brown},
  {Vallenari}, {Prusti}, {de Bruijne}, {Babusiaux}, {Biermann}, {Creevey},
  {Evans}, {Eyer}, {Hutton}, {Jansen}, {Jordi}, {Klioner}, {Lammers},
  {Lindegren}, {Luri}, {Mignard}, {Panem}, {Pourbaix}, {Randich}, {Sartoretti},
  {Soubiran}, {Walton}, {Arenou}, {Bailer-Jones}, {Bastian}, {Cropper},
  {Drimmel}, {Katz}, {Lattanzi}, {van Leeuwen}, {Bakker}, {Cacciari},
  {Casta{\~n}eda}, {De Angeli}, {Ducourant}, {Fabricius}, {Fouesneau},
  {Fr{\'e}mat}, {Guerra}, {Guerrier}, {Guiraud}, {Jean-Antoine Piccolo},
  {Masana}, {Messineo}, {Mowlavi}, {Nicolas}, {Nienartowicz}, {Pailler},
  {Panuzzo}, {Riclet}, {Roux}, {Seabroke}, {Sordo}, {Tanga}, {Th{\'e}venin},
  {Gracia-Abril}, {Portell}, {Teyssier}, {Altmann}, {Andrae}, {Bellas-Velidis},
  {Benson}, {Berthier}, {Blomme}, {Brugaletta}, {Burgess}, {Busso}, {Carry},
  {Cellino}, {Cheek}, {Clementini}, {Damerdji}, {Davidson}, {Delchambre},
  {Dell'Oro}, {Fern{\'a}ndez-Hern{\'a}ndez}, {Galluccio}, {Garc{\'\i}a-Lario},
  {Garcia-Reinaldos}, {Gonz{\'a}lez-N{\'u}{\~n}ez}, {Gosset}, {Haigron},
  {Halbwachs}, {Hambly}, {Harrison}, {Hatzidimitriou}, {Heiter},
  {Hern{\'a}ndez}, {Hestroffer}, {Hodgkin}, {Holl}, {Jan{\ss}en}, {Jevardat de
  Fombelle}, {Jordan}, {Krone-Martins}, {Lanzafame}, {L{\"o}ffler}, {Lorca},
  {Manteiga}, {Marchal}, {Marrese}, {Moitinho}, {Mora}, {Muinonen}, {Osborne},
  {Pancino}, {Pauwels}, {Petit}, {Recio-Blanco}, {Richards}, {Riello},
  {Rimoldini}, {Robin}, {Roegiers}, {Rybizki}, {Sarro}, {Siopis}, {Smith},
  {Sozzetti}, {Ulla}, {Utrilla}, {van Leeuwen}, {van Reeven}, {Abbas}, {Abreu
  Aramburu}, {Accart}, {Aerts}, {Aguado}, {Ajaj}, {Altavilla}, {{\'A}lvarez},
  {{\'A}lvarez Cid-Fuentes}, {Alves}, {Anderson}, {Anglada Varela}, {Antoja},
  {Audard}, {Baines}, {Baker}, {Balaguer-N{\'u}{\~n}ez}, {Balbinot}, {Balog},
  {Barache}, {Barbato}, {Barros}, {Barstow}, {Bartolom{\'e}}, {Bassilana},
  {Bauchet}, {Baudesson-Stella}, {Becciani}, {Bellazzini}, {Bernet}, {Bertone},
  {Bianchi}, {Blanco-Cuaresma}, {Boch}, {Bombrun}, {Bossini}, {Bouquillon},
  {Bragaglia}, {Bramante}, {Breedt}, {Bressan}, {Brouillet}, {Bucciarelli},
  {Burlacu}, {Busonero}, {Butkevich}, {Buzzi}, {Caffau}, {Cancelliere},
  {C{\'a}novas}, {Cantat-Gaudin}, {Carballo}, {Carlucci}, {Carnerero},
  {Carrasco}, {Casamiquela}, {Castellani}, {Castro-Ginard}, {Castro Sampol},
  {Chaoul}, {Charlot}, {Chemin}, {Chiavassa}, {Cioni}, {Comoretto}, {Cooper},
  {Cornez}, {Cowell}, {Crifo}, {Crosta}, {Crowley}, {Dafonte}, {Dapergolas},
  {David}, {David}, {de Laverny}, {De Luise}, {De March}, {De Ridder}, {de
  Souza}, {de Teodoro}, {de Torres}, {del Peloso}, {del Pozo}, {Delbo},
  {Delgado}, {Delgado}, {Delisle}, {Di Matteo}, {Diakite}, {Diener},
  {Distefano}, {Dolding}, {Eappachen}, {Edvardsson}, {Enke}, {Esquej}, {Fabre},
  {Fabrizio}, {Faigler}, {Fedorets}, {Fernique}, {Fienga}, {Figueras},
  {Fouron}, {Fragkoudi}, {Fraile}, {Franke}, {Gai}, {Garabato},
  {Garcia-Gutierrez}, {Garc{\'\i}a-Torres}, {Garofalo}, {Gavras}, {Gerlach},
  {Geyer}, {Giacobbe}, {Gilmore}, {Girona}, {Giuffrida}, {Gomel}, {Gomez},
  {Gonzalez-Santamaria}, {Gonz{\'a}lez-Vidal}, {Granvik},
  {Guti{\'e}rrez-S{\'a}nchez}, {Guy}, {Hauser}, {Haywood}, {Helmi}, {Hidalgo},
  {Hilger}, {H{\l}adczuk}, {Hobbs}, {Holland}, {Huckle}, {Jasniewicz},
  {Jonker}, {Juaristi Campillo}, {Julbe}, {Karbevska}, {Kervella}, {Khanna},
  {Kochoska}, {Kontizas}, {Kordopatis}, {Korn}, {Kostrzewa-Rutkowska},
  {Kruszy{\'n}ska}, {Lambert}, {Lanza}, {Lasne}, {Le Campion}, {Le Fustec},
  {Lebreton}, {Lebzelter}, {Leccia}, {Leclerc}, {Lecoeur-Taibi}, {Liao},
  {Licata}, {Lindstr{\o}m}, {Lister}, {Livanou}, {Lobel}, {Madrero Pardo},
  {Managau}, {Mann}, {Marchant}, {Marconi}, {Marcos Santos}, {Marinoni},
  {Marocco}, {Marshall}, {Martin Polo}, {Mart{\'\i}n-Fleitas}, {Masip},
  {Massari}, {Mastrobuono-Battisti}, {Mazeh}, {McMillan}, {Messina},
  {Michalik}, {Millar}, {Mints}, {Molina}, {Molinaro}, {Moln{\'a}r},
  {Montegriffo}, {Mor}, {Morbidelli}, {Morel}, {Morris}, {Mulone}, {Munoz},
  {Muraveva}, {Murphy}, {Musella}, {Noval}, {Ord{\'e}novic}, {Orr{\`u}},
  {Osinde}, {Pagani}, {Pagano}, {Palaversa}, {Palicio}, {Panahi}, {Pawlak},
  {Pe{\~n}alosa Esteller}, {Penttil{\"a}}, {Piersimoni}, {Pineau}, {Plachy},
  {Plum}, {Poggio}, {Poretti}, {Poujoulet}, {Pr{\v{s}}a}, {Pulone}, {Racero},
  {Ragaini}, {Rainer}, {Raiteri}, {Rambaux}, {Ramos}, {Ramos-Lerate}, {Re
  Fiorentin}, {Regibo}, {Reyl{\'e}}, {Ripepi}, {Riva}, {Rixon}, {Robichon},
  {Robin}, {Roelens}, {Rohrbasser}, {Romero-G{\'o}mez}, {Rowell}, {Royer},
  {Rybicki}, {Sadowski}, {Sagrist{\`a} Sell{\'e}s}, {Sahlmann}, {Salgado},
  {Salguero}, {Samaras}, {Sanchez Gimenez}, {Sanna}, {Santove{\~n}a},
  {Sarasso}, {Schultheis}, {Sciacca}, {Segol}, {Segovia}, {S{\'e}gransan},
  {Semeux}, {Shahaf}, {Siddiqui}, {Siebert}, {Siltala}, {Slezak}, {Smart},
  {Solano}, {Solitro}, {Souami}, {Souchay}, {Spagna}, {Spoto}, {Steele},
  {Steidelm{\"u}ller}, {Stephenson}, {S{\"u}veges}, {Szabados}, {Szegedi-Elek},
  {Taris}, {Tauran}, {Taylor}, {Teixeira}, {Thuillot}, {Tonello}, {Torra},
  {Torra}, {Turon}, {Unger}, {Vaillant}, {van Dillen}, {Vanel}, {Vecchiato},
  {Viala}, {Vicente}, {Voutsinas}, {Weiler}, {Wevers}, {Wyrzykowski}, {Yoldas},
  {Yvard}, {Zhao}, {Zorec}, {Zucker}, {Zurbach}, \&
  {Zwitter}}]{gaia2021A&A...649A...1G}
{Gaia Collaboration}, {Brown}, A.~G.~A., {Vallenari}, A., {et~al.} 2021, \aap,
  649, A1

\bibitem[{{Girichidis} {et~al.}(2020){Girichidis}, {Offner}, {Kritsuk},
  {Klessen}, {Hennebelle}, {Kruijssen}, {Krause}, {Glover}, \&
  {Padovani}}]{girichidis2020}
{Girichidis}, P., {Offner}, S. S.~R., {Kritsuk}, A.~G., {et~al.} 2020, \ssr,
  216, 68

\bibitem[{{Gro{\ss}schedl} {et~al.}(2021){Gro{\ss}schedl}, {Alves}, {Meingast},
  \& {Herbst-Kiss}}]{grossschedl2021}
{Gro{\ss}schedl}, J.~E., {Alves}, J., {Meingast}, S., \& {Herbst-Kiss}, G.
  2021, \aap, 647, A91

\bibitem[{{Guesten} \& {Mezger}(1982)}]{guestenANDmezger1982}
{Guesten}, R. \& {Mezger}, P.~G. 1982, Vistas in Astronomy, 26, 159

\bibitem[{{Heyer} \& {Dame}(2015)}]{heyerANDdame2015}
{Heyer}, M. \& {Dame}, T.~M. 2015, \araa, 53, 583

\bibitem[{{Hunter} {et~al.}(2023){Hunter}, {Klessen}, {Sormani}, {Glover},
  {Soler}, {Molinari}, {Hennebelle}, \& {Testi}}]{hunter2023}
{Hunter}, G.~H., {Klessen}, R.~S., {Sormani}, M.~C., {et~al.} 2023, In prep.

\bibitem[{{Kennicutt}(1989)}]{kennicutt1989}
{Kennicutt}, Robert~C., J. 1989, \apj, 344, 685

\bibitem[{{Kennicutt} \& {Evans}(2012)}]{kennicuttANDevans2012}
{Kennicutt}, R.~C. \& {Evans}, N.~J. 2012, \araa, 50, 531

\bibitem[{{Kim} {et~al.}(2022){Kim}, {Chevance}, {Kruijssen}, {Leroy},
  {Schruba}, {Barnes}, {Bigiel}, {Blanc}, {Cao}, {Congiu}, {Dale}, {Faesi},
  {Glover}, {Grasha}, {Groves}, {Hughes}, {Klessen}, {Kreckel}, {McElroy},
  {Pan}, {Pety}, {Querejeta}, {Razza}, {Rosolowsky}, {Saito}, {Schinnerer},
  {Sun}, {Tomi{\v{c}}i{\'c}}, {Usero}, \& {Williams}}]{kim2022}
{Kim}, J., {Chevance}, M., {Kruijssen}, J.~M.~D., {et~al.} 2022, \mnras, 516,
  3006

\bibitem[{{Klessen} \& {Glover}(2016)}]{klessenANDglover2016}
{Klessen}, R.~S. \& {Glover}, S. C.~O. 2016, in Saas-Fee Advanced Course,
  Vol.~43, Saas-Fee Advanced Course, ed. Y.~{Revaz}, P.~{Jablonka},
  R.~{Teyssier}, \& L.~{Mayer}, 85

\bibitem[{{K{\"o}nyves} {et~al.}(2015){K{\"o}nyves}, {Andr{\'e}},
  {Men'shchikov}, {Palmeirim}, {Arzoumanian}, {Schneider}, {Roy}, {Didelon},
  {Maury}, {Shimajiri}, {Di Francesco}, {Bontemps}, {Peretto}, {Benedettini},
  {Bernard}, {Elia}, {Griffin}, {Hill}, {Kirk}, {Ladjelate}, {Marsh}, {Martin},
  {Motte}, {Nguy{\^e}n Luong}, {Pezzuto}, {Roussel}, {Rygl}, {Sadavoy},
  {Schisano}, {Spinoglio}, {Ward-Thompson}, \& {White}}]{konyves2015}
{K{\"o}nyves}, V., {Andr{\'e}}, P., {Men'shchikov}, A., {et~al.} 2015, \aap,
  584, A91

\bibitem[{{Kreckel} {et~al.}(2018){Kreckel}, {Faesi}, {Kruijssen}, {Schruba},
  {Groves}, {Leroy}, {Bigiel}, {Blanc}, {Chevance}, {Herrera}, {Hughes},
  {McElroy}, {Pety}, {Querejeta}, {Rosolowsky}, {Schinnerer}, {Sun}, {Usero},
  \& {Utomo}}]{kreckel2018}
{Kreckel}, K., {Faesi}, C., {Kruijssen}, J.~M.~D., {et~al.} 2018, \apjl, 863,
  L21

\bibitem[{{Kroupa}(2002)}]{kroupa2002}
{Kroupa}, P. 2002, Science, 295, 82

\bibitem[{{Kroupa} \& {Weidner}(2003)}]{Kroupa2003}
{Kroupa}, P. \& {Weidner}, C. 2003, \apj, 598, 1076

\bibitem[{{Krumholz} {et~al.}(2014){Krumholz}, {Bate}, {Arce}, {Dale},
  {Gutermuth}, {Klein}, {Li}, {Nakamura}, \& {Zhang}}]{krumholz2014}
{Krumholz}, M.~R., {Bate}, M.~R., {Arce}, H.~G., {et~al.} 2014, in Protostars
  and Planets VI, ed. H.~{Beuther}, R.~S. {Klessen}, C.~P. {Dullemond}, \&
  T.~{Henning}, 243--266

\bibitem[{{Lada} {et~al.}(2010){Lada}, {Lombardi}, \& {Alves}}]{lada2010}
{Lada}, C.~J., {Lombardi}, M., \& {Alves}, J.~F. 2010, \apj, 724, 687

\bibitem[{{Lee} {et~al.}(2020){Lee}, {Offner}, {Hennebelle}, {Andr{\'e}},
  {Zinnecker}, {Ballesteros-Paredes}, {Inutsuka}, \& {Kruijssen}}]{lee2020}
{Lee}, Y.-N., {Offner}, S. S.~R., {Hennebelle}, P., {et~al.} 2020, \ssr, 216,
  70

\bibitem[{{Leroy} {et~al.}(2013){Leroy}, {Walter}, {Sandstrom}, {Schruba},
  {Munoz-Mateos}, {Bigiel}, {Bolatto}, {Brinks}, {de Blok}, {Meidt}, {Rix},
  {Rosolowsky}, {Schinnerer}, {Schuster}, \& {Usero}}]{leroy2013}
{Leroy}, A.~K., {Walter}, F., {Sandstrom}, K., {et~al.} 2013, \aj, 146, 19

\bibitem[{{Licquia} \& {Newman}(2015)}]{licquiaANDnewman2015}
{Licquia}, T.~C. \& {Newman}, J.~A. 2015, \apj, 806, 96

\bibitem[{{McKee} \& {Ostriker}(2007)}]{mckeeANDostriker2007}
{McKee}, C.~F. \& {Ostriker}, E.~C. 2007, \araa, 45, 565

\bibitem[{{McKee} \& {Williams}(1997)}]{mckeeANDwilliams1997}
{McKee}, C.~F. \& {Williams}, J.~P. 1997, \apj, 476, 144

\bibitem[{{M{\`e}ge} {et~al.}(2021){M{\`e}ge}, {Russeil}, {Zavagno}, {Elia},
  {Molinari}, {Brunt}, {Butora}, {Cambresy}, {Di Giorgio}, {Fenouillet},
  {Fukui}, {Lambert}, {Makai}, {Merello}, {Meunier}, {Molinaro}, {Moreau},
  {Pezzuto}, {Poulin}, {Schisano}, \& {Schuller}}]{mege2021}
{M{\`e}ge}, P., {Russeil}, D., {Zavagno}, A., {et~al.} 2021, \aap, 646, A74

\bibitem[{{Megeath} {et~al.}(2008){Megeath}, {Townsley}, {Oey}, \&
  {Tieftrunk}}]{megeath2008}
{Megeath}, S.~T., {Townsley}, L.~K., {Oey}, M.~S., \& {Tieftrunk}, A.~R. 2008,
  in Handbook of Star Forming Regions, Volume I, ed. B.~{Reipurth}, Vol.~4, 264

\bibitem[{{Miville-Desch{\^e}nes} {et~al.}(2017){Miville-Desch{\^e}nes},
  {Murray}, \& {Lee}}]{miville-deschenes2017}
{Miville-Desch{\^e}nes}, M.-A., {Murray}, N., \& {Lee}, E.~J. 2017, \apj, 834,
  57

\bibitem[{{Molinari} {et~al.}(2008){Molinari}, {Pezzuto}, {Cesaroni}, {Brand},
  {Faustini}, \& {Testi}}]{molinari2008}
{Molinari}, S., {Pezzuto}, S., {Cesaroni}, R., {et~al.} 2008, \aap, 481, 345

\bibitem[{{Molinari} {et~al.}(2016){Molinari}, {Schisano}, {Elia},
  {Pestalozzi}, {Traficante}, {Pezzuto}, {Swinyard}, {Noriega-Crespo}, {Bally},
  {Moore}, {Plume}, {Zavagno}, {di Giorgio}, {Liu}, {Pilbratt}, {Mottram},
  {Russeil}, {Piazzo}, {Veneziani}, {Benedettini}, {Calzoletti}, {Faustini},
  {Natoli}, {Piacentini}, {Merello}, {Palmese}, {Del Grande}, {Polychroni},
  {Rygl}, {Polenta}, {Barlow}, {Bernard}, {Martin}, {Testi}, {Ali},
  {Andr{\'e}}, {Beltr{\'a}n}, {Billot}, {Carey}, {Cesaroni}, {Compi{\`e}gne},
  {Eden}, {Fukui}, {Garcia-Lario}, {Hoare}, {Huang}, {Joncas}, {Lim}, {Lord},
  {Martinavarro-Armengol}, {Motte}, {Paladini}, {Paradis}, {Peretto},
  {Robitaille}, {Schilke}, {Schneider}, {Schulz}, {Sibthorpe}, {Strafella},
  {Thompson}, {Umana}, {Ward-Thompson}, \& {Wyrowski}}]{molinari2016}
{Molinari}, S., {Schisano}, E., {Elia}, D., {et~al.} 2016, \aap, 591, A149

\bibitem[{{Murray} \& {Rahman}(2010)}]{murrayANDrahman2010}
{Murray}, N. \& {Rahman}, M. 2010, \apj, 709, 424

\bibitem[{{Ossenkopf} \& {Henning}(1994)}]{ossenkopf1994}
{Ossenkopf}, V. \& {Henning}, T. 1994, \aap, 291, 943

\bibitem[{{Paradis} {et~al.}(2010){Paradis}, {Veneziani}, {Noriega-Crespo},
  {Paladini}, {Piacentini}, {Bernard}, {de Bernardis}, {Calzoletti},
  {Faustini}, {Martin}, {Masi}, {Montier}, {Natoli}, {Ristorcelli}, {Thompson},
  {Traficante}, \& {Molinari}}]{paradis2010}
{Paradis}, D., {Veneziani}, M., {Noriega-Crespo}, A., {et~al.} 2010, \aap, 520,
  L8

\bibitem[{{Peek} {et~al.}(2022){Peek}, {Tchernyshyov}, \&
  {Miville-Deschenes}}]{peek2022}
{Peek}, J.~E.~G., {Tchernyshyov}, K., \& {Miville-Deschenes}, M.-A. 2022, \apj,
  925, 201

\bibitem[{{Planck Collaboration XI.}(2014)}]{planck2013-p06b}
{Planck Collaboration XI.} 2014, \aap, 571, A11

\bibitem[{{Reipurth}(2008{\natexlab{a}})}]{reipurth2008a}
{Reipurth}, B. 2008{\natexlab{a}}, {Handbook of Star Forming Regions, Volume I:
  The Northern Sky}, Vol.~4

\bibitem[{{Reipurth}(2008{\natexlab{b}})}]{reipurth2008b}
{Reipurth}, B. 2008{\natexlab{b}}, {Handbook of Star Forming Regions, Volume
  II: The Southern Sky}, Vol.~5

\bibitem[{{Reipurth} \& {Schneider}(2008)}]{reipurth2008cygnus}
{Reipurth}, B. \& {Schneider}, N. 2008, in Handbook of Star Forming Regions,
  Volume I, ed. B.~{Reipurth}, Vol.~4, 36

\bibitem[{{Riener} {et~al.}(2020){Riener}, {Kainulainen}, {Henshaw}, \&
  {Beuther}}]{riener2020}
{Riener}, M., {Kainulainen}, J., {Henshaw}, J.~D., \& {Beuther}, H. 2020, \aap,
  640, A72

\bibitem[{{Robitaille} \& {Whitney}(2010)}]{robitaille2010}
{Robitaille}, T.~P. \& {Whitney}, B.~A. 2010, \apjl, 710, L11

\bibitem[{{Sadavoy} {et~al.}(2013){Sadavoy}, {Di Francesco}, {Johnstone},
  {Currie}, {Drabek}, {Hatchell}, {Nutter}, {Andr{\'e}}, {Arzoumanian},
  {Benedettini}, {Bernard}, {Duarte-Cabral}, {Fallscheer}, {Friesen},
  {Greaves}, {Hennemann}, {Hill}, {Jenness}, {K{\"o}nyves}, {Matthews},
  {Mottram}, {Pezzuto}, {Roy}, {Rygl}, {Schneider-Bontemps}, {Spinoglio},
  {Testi}, {Tothill}, {Ward-Thompson}, {White}, {JCMT}, \& {Herschel Gould Belt
  Survey Teams}}]{sadavoy2013}
{Sadavoy}, S.~I., {Di Francesco}, J., {Johnstone}, D., {et~al.} 2013, \apj,
  767, 126

\bibitem[{{Salpeter}(1955)}]{salpeter1955}
{Salpeter}, E.~E. 1955, \apj, 121, 161

\bibitem[{Saxe(1868)}]{saxe1868poems}
Saxe, J. 1868, The Poems of John Godfrey Saxe: Complete in One Volume (Ticknor
  and Fields)

\bibitem[{{Schruba} {et~al.}(2011){Schruba}, {Leroy}, {Walter}, {Bigiel},
  {Brinks}, {de Blok}, {Dumas}, {Kramer}, {Rosolowsky}, {Sandstrom},
  {Schuster}, {Usero}, {Weiss}, \& {Wiesemeyer}}]{schruba2011}
{Schruba}, A., {Leroy}, A.~K., {Walter}, F., {et~al.} 2011, \aj, 142, 37

\bibitem[{{Schuller} {et~al.}(2009){Schuller}, {Menten}, {Contreras},
  {Wyrowski}, {Schilke}, {Bronfman}, {Henning}, {Walmsley}, {Beuther},
  {Bontemps}, {Cesaroni}, {Deharveng}, {Garay}, {Herpin}, {Lefloch}, {Linz},
  {Mardones}, {Minier}, {Molinari}, {Motte}, {Nyman}, {Reveret}, {Risacher},
  {Russeil}, {Schneider}, {Testi}, {Troost}, {Vasyunina}, {Wienen}, {Zavagno},
  {Kovacs}, {Kreysa}, {Siringo}, \& {Wei{\ss}}}]{schuller2009}
{Schuller}, F., {Menten}, K.~M., {Contreras}, Y., {et~al.} 2009, \aap, 504, 415

\bibitem[{{Skrutskie} {et~al.}(2006){Skrutskie}, {Cutri}, {Stiening},
  {Weinberg}, {Schneider}, {Carpenter}, {Beichman}, {Capps}, {Chester},
  {Elias}, {Huchra}, {Liebert}, {Lonsdale}, {Monet}, {Price}, {Seitzer},
  {Jarrett}, {Kirkpatrick}, {Gizis}, {Howard}, {Evans}, {Fowler}, {Fullmer},
  {Hurt}, {Light}, {Kopan}, {Marsh}, {McCallon}, {Tam}, {Van Dyk}, \&
  {Wheelock}}]{skrutskie2006}
{Skrutskie}, M.~F., {Cutri}, R.~M., {Stiening}, R., {et~al.} 2006, \aj, 131,
  1163

\bibitem[{{Smith} {et~al.}(1978){Smith}, {Biermann}, \& {Mezger}}]{smith1978}
{Smith}, L.~F., {Biermann}, P., \& {Mezger}, P.~G. 1978, \aap, 66, 65

\bibitem[{{Smith} \& {Brooks}(2008)}]{smith2008}
{Smith}, N. \& {Brooks}, K.~J. 2008, in Handbook of Star Forming Regions,
  Volume II, ed. B.~{Reipurth}, Vol.~5, 138

\bibitem[{{Sun} {et~al.}(2023){Sun}, {Leroy}, {Ostriker}, {Meidt},
  {Rosolowsky}, {Schinnerer}, {Wilson}, {Utomo}, {Belfiore}, {Blanc},
  {Emsellem}, {Faesi}, {Groves}, {Hughes}, {Koch}, {Kreckel}, {Liu}, {Pan},
  {Pety}, {Querejeta}, {Razza}, {Saito}, {Sardone}, {Usero}, {Williams},
  {Bigiel}, {Bolatto}, {Chevance}, {Dale}, {Gensior}, {Glover}, {Grasha},
  {Henshaw}, {Jim{\'e}nez-Donaire}, {Klessen}, {Kruijssen}, {Murphy},
  {Neumann}, {Teng}, \& {Thilker}}]{sun2023}
{Sun}, J., {Leroy}, A.~K., {Ostriker}, E.~C., {et~al.} 2023, \apjl, 945, L19

\bibitem[{{Tacconi} {et~al.}(2020){Tacconi}, {Genzel}, \&
  {Sternberg}}]{Tacconi2020}
{Tacconi}, L.~J., {Genzel}, R., \& {Sternberg}, A. 2020, \araa, 58, 157

\bibitem[{{Urquhart} {et~al.}(2022){Urquhart}, {Wells}, {Pillai}, {Leurini},
  {Giannetti}, {Moore}, {Thompson}, {Figura}, {Colombo}, {Yang}, {K{\"o}nig},
  {Wyrowski}, {Menten}, {Rigby}, {Eden}, \& {Ragan}}]{urquhart2022}
{Urquhart}, J.~S., {Wells}, M.~R.~A., {Pillai}, T., {et~al.} 2022, \mnras, 510,
  3389

\bibitem[{{Veneziani} {et~al.}(2013){Veneziani}, {Elia}, {Noriega-Crespo},
  {Paladini}, {Carey}, {Faimali}, {Molinari}, {Pestalozzi}, {Piacentini},
  {Schisano}, \& {Tibbs}}]{veneziani2013}
{Veneziani}, M., {Elia}, D., {Noriega-Crespo}, A., {et~al.} 2013, \aap, 549,
  A130

\bibitem[{{Veneziani} {et~al.}(2017){Veneziani}, {Schisano}, {Elia},
  {Noriega-Crespo}, {Carey}, {Di Giorgio}, {Fukui}, {Maiolo}, {Maruccia},
  {Mizuno}, {Mizuno}, {Molinari}, {Mottram}, {Moore}, {Onishi}, {Paladini},
  {Paradis}, {Pestalozzi}, {Pezzuto}, {Piacentini}, {Plume}, {Russeil}, \&
  {Strafella}}]{veneziani2017}
{Veneziani}, M., {Schisano}, E., {Elia}, D., {et~al.} 2017, \aap, 599, A7

\bibitem[{{Weaver} {et~al.}(1977){Weaver}, {McCray}, {Castor}, {Shapiro}, \&
  {Moore}}]{weaver1977}
{Weaver}, R., {McCray}, R., {Castor}, J., {Shapiro}, P., \& {Moore}, R. 1977,
  \apj, 218, 377

\bibitem[{{Wells} {et~al.}(2022){Wells}, {Urquhart}, {Moore}, {Browning},
  {Ragan}, {Rigby}, {Eden}, \& {Thompson}}]{wells2022}
{Wells}, M.~R.~A., {Urquhart}, J.~S., {Moore}, T.~J.~T., {et~al.} 2022, \mnras,
  516, 4245

\bibitem[{{Wenger} {et~al.}(2018){Wenger}, {Balser}, {Anderson}, \&
  {Bania}}]{wegner2018}
{Wenger}, T.~V., {Balser}, D.~S., {Anderson}, L.~D., \& {Bania}, T.~M. 2018,
  \apj, 856, 52

\bibitem[{{Wolk} {et~al.}(2008){Wolk}, {Bourke}, \& {Vigil}}]{wolk2008}
{Wolk}, S.~J., {Bourke}, T.~L., \& {Vigil}, M. 2008, in Handbook of Star
  Forming Regions, Volume II, ed. B.~{Reipurth}, Vol.~5, 124

\bibitem[{{Zari} {et~al.}(2023){Zari}, {Frankel}, \& {Rix}}]{zari2023}
{Zari}, E., {Frankel}, N., \& {Rix}, H.-W. 2023, \aap, 669, A10

\bibitem[{{Zari} {et~al.}(2021){Zari}, {Rix}, {Frankel}, {Xiang}, {Poggio},
  {Drimmel}, \& {Tkachenko}}]{zari2021}
{Zari}, E., {Rix}, H.~W., {Frankel}, N., {et~al.} 2021, \aap, 650, A112

\bibitem[{{Zucker} {et~al.}(2020){Zucker}, {Speagle}, {Schlafly}, {Green},
  {Finkbeiner}, {Goodman}, \& {Alves}}]{zucker2020}
{Zucker}, C., {Speagle}, J.~S., {Schlafly}, E.~F., {et~al.} 2020, \aap, 633,
  A51

\end{thebibliography}

% ===============================================================================================
% APPENDICES APENDICES 
% =======================================================================================
\appendix

% ================================================================
\section{Construction of radial profiles}\label{appendix:method}

\begin{figure}[ht!]
\centerline{
\includegraphics[width=0.499\textwidth,angle=0,origin=c]{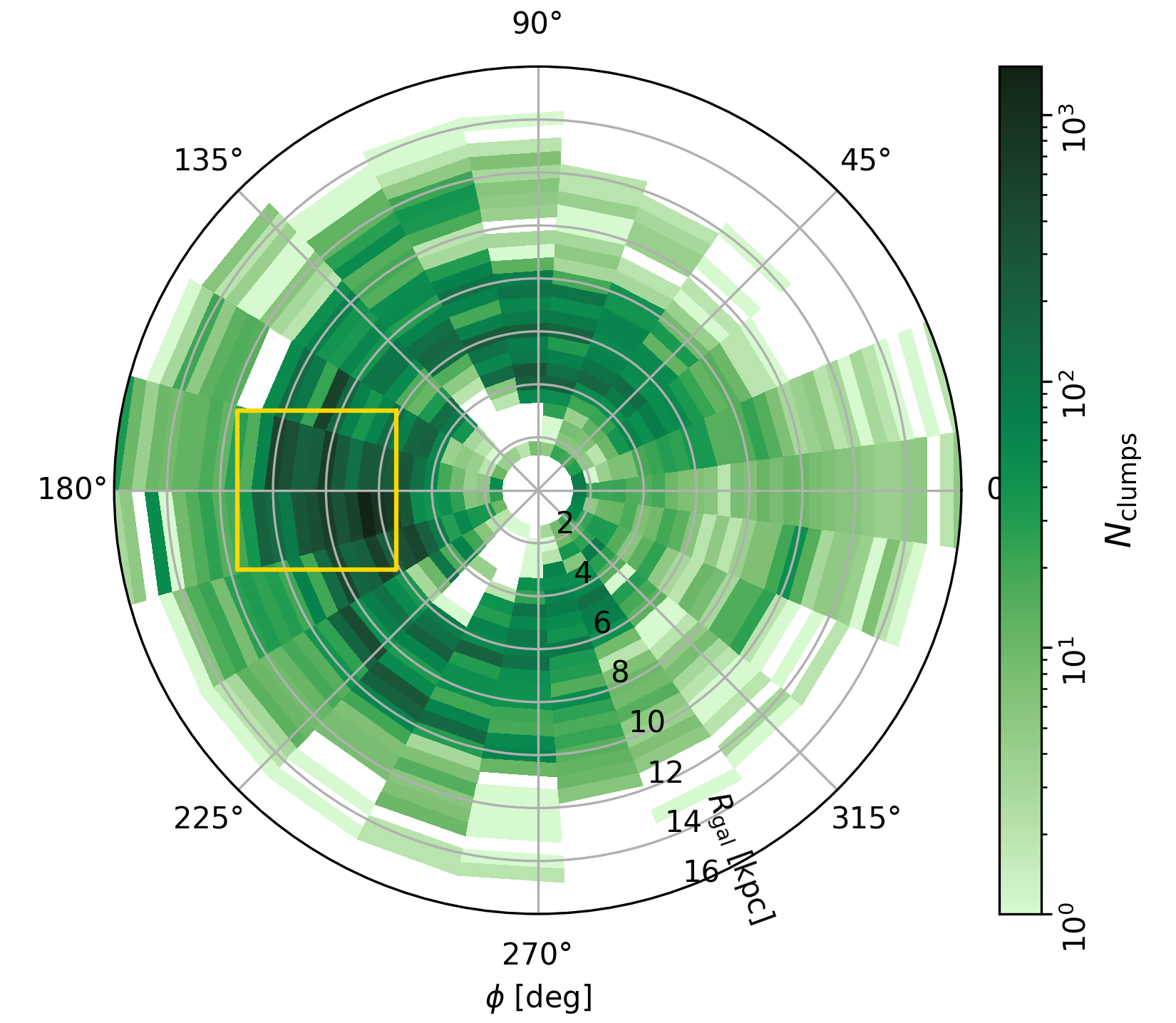}
}
\centerline{
\includegraphics[width=0.499\textwidth,angle=0,origin=c]{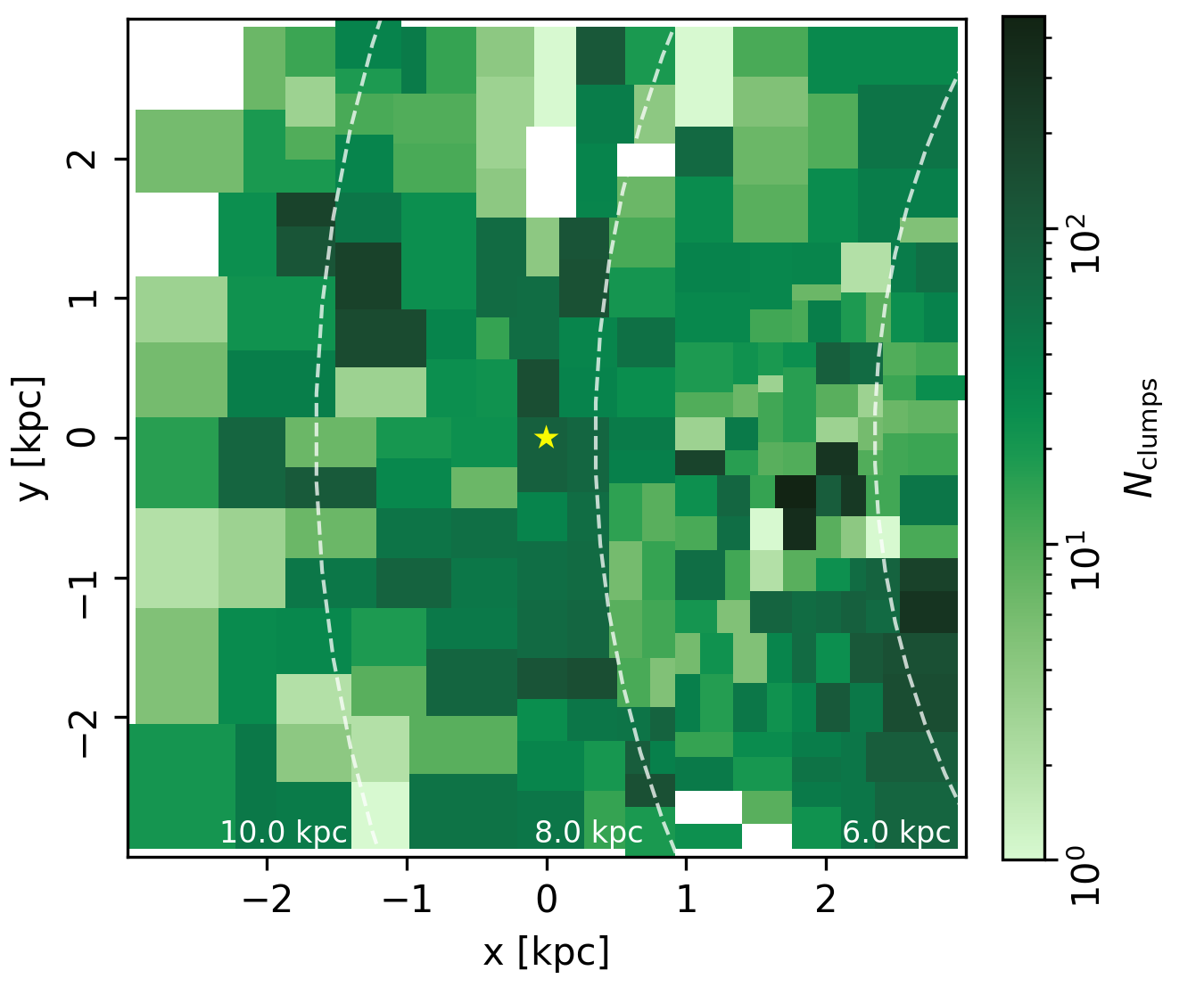}
}
\caption{Number of \HIGAL\ clumps in the polar and Cartesian grids presented in Figs.~\ref{fig:SFRpolargrid} and \ref{fig:SFRmaps}.}
\label{fig:NclumpsMap}
\end{figure}

\begin{figure}[ht!]
\centerline{
\includegraphics[width=0.499\textwidth,angle=0,origin=c]{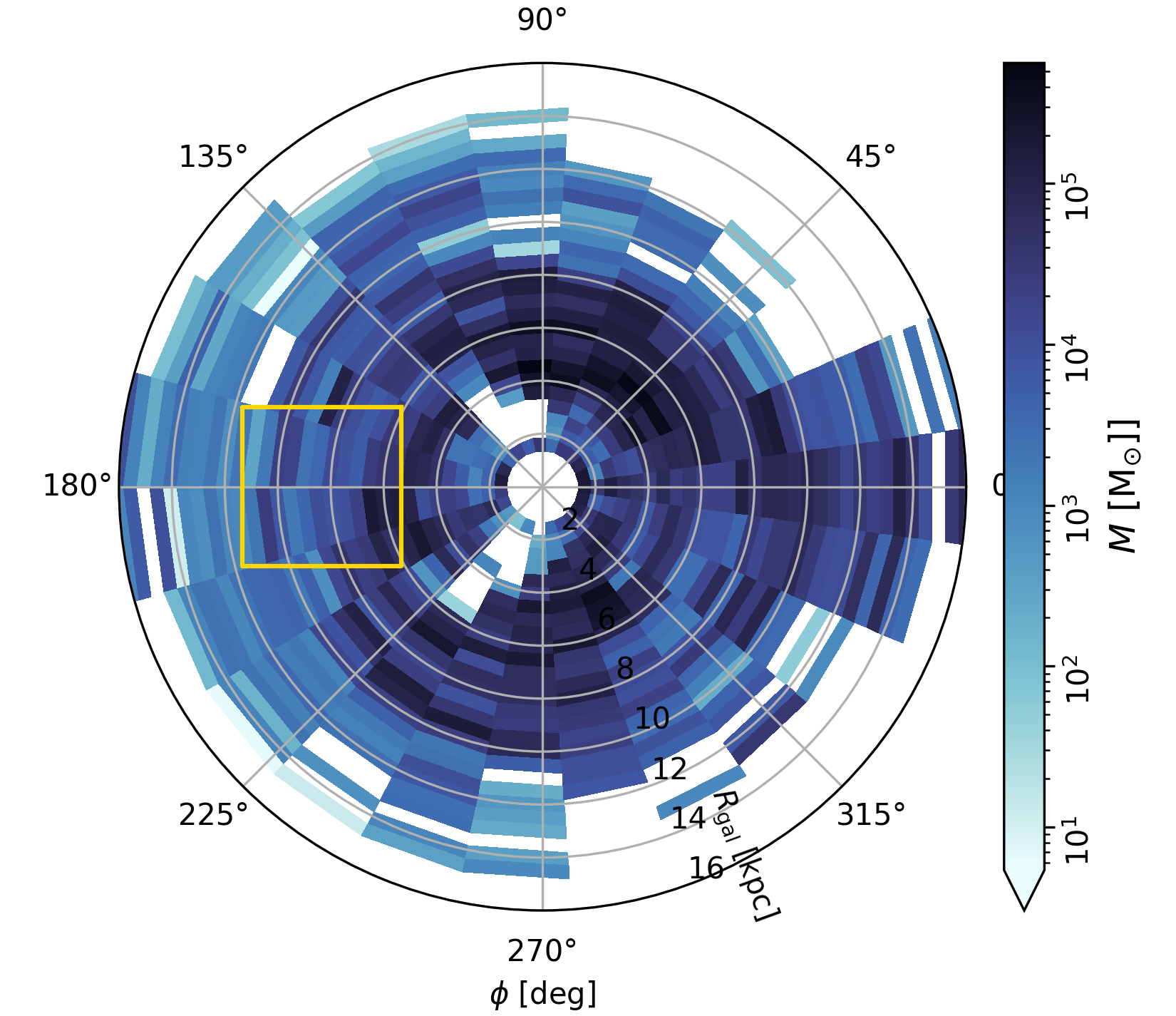}
}
\centerline{
\includegraphics[width=0.499\textwidth,angle=0,origin=c]{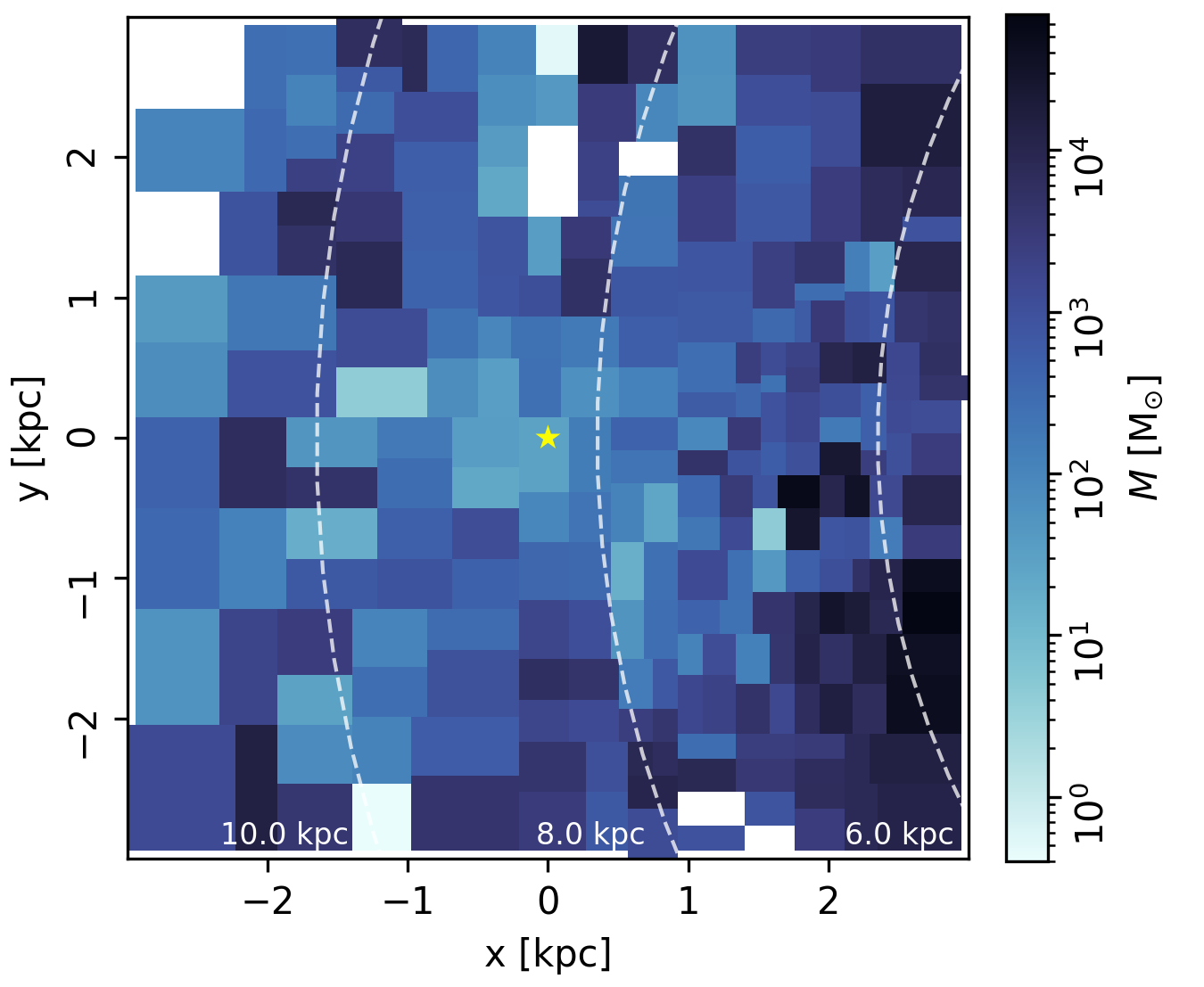}
}
\caption{Total mass in \HIGAL\ clumps in the polar and Cartesian grids presented in Figs.~\ref{fig:SFRpolargrid} and \ref{fig:SFRmaps}.}
\label{fig:MassMap}
\end{figure}

In this appendix we present some of the quantities involved in the calculation of the $\Sigma_{\rm SFR}$ radial profiles and maps presented in the main body of this paper.
Figure~\ref{fig:NclumpsMap} shows the number of E22 clumps distributed in the polar grid presented in Fig.~\ref{fig:SFRpolargrid} and the maps in Fig.~\ref{fig:SFRmaps}.
We computed \SigmaSFR\ by adding the contributions derived from Eq.~\ref{eq:mass2sfr} for the clumps in each spaxel.
For comparison, we present in Fig.~\ref{fig:MassMap} the sum of the clump masses in each spaxel for the Galactic polar and Z23 grids.

% =============================================
\section{Effect on the IMF selection}\label{appendix:IMF}

\begin{figure}[ht!]
\centerline{\includegraphics[width=0.49\textwidth,angle=0,origin=c]{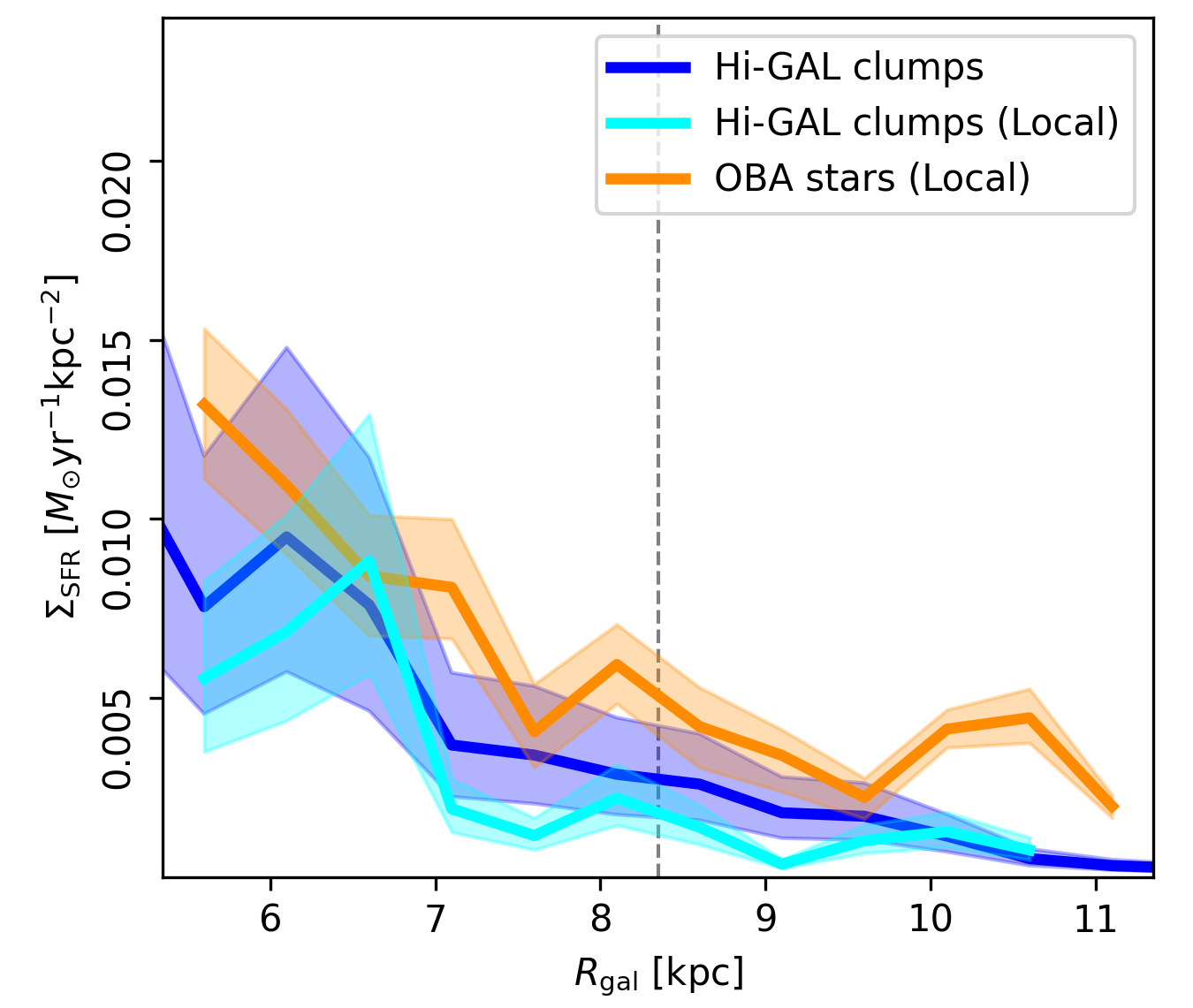}}
\caption{Same as Fig.~\ref{fig:SFRprofiles}, but for a two-component broken-power-law IMF.
}
\label{fig:SFRprofilesIMF2}
\end{figure}

We considered the results of an alternative IMF selection in the modeling of the high-mass stellar population by employing a two-component broken power law with $\Gamma$\,$=$\,$1.3$ for masses between 0.1 and 0.5\,$M_{\odot}$ and $\Gamma$\,$=$\,$2.3$ for masses between 0.5 and 100\,$M_{\odot}$ \citep{Kroupa2003}.
The results, presented in Fig.~\ref{fig:SFRprofilesIMF2}, are much closer to the E22 \SigmaSFR\ profile.
However, this selection of IMF does not account for unresolved binaries, which comprise a large portion of the objects in the Z23 source catalog.

\end{document}